\documentclass[sn-basic,iicol]{sn-jnl}
\usepackage{graphicx}
\usepackage{multirow}
\usepackage{amsthm}
\usepackage{mathrsfs}
\usepackage[title]{appendix}
\usepackage{manyfoot}
\usepackage{booktabs}
\usepackage{listings}
\usepackage[most]{tcolorbox}

\usepackage{amsmath,amssymb,amsfonts}
\usepackage[ruled,vlined]{algorithm2e}
\usepackage[inline]{enumitem}
\usepackage{graphicx}
\usepackage{textcomp}
\usepackage{soul, color}
\usepackage{tabularx}
\usepackage{hyperref}
\usepackage{mathtools}
\usepackage{booktabs}
\usepackage{bookmark}
\usepackage{threeparttable}
\usepackage{todonotes}
\usepackage{rotating}
\usepackage{lmodern}
\usepackage{anyfontsize}

\soulregister\cite7
\soulregister\ref7
\soulregister\pageref7

\hypersetup{
  breaklinks   = true, 
  colorlinks   = true,    
  urlcolor     = blue,    
  linkcolor    = blue,    
  citecolor    = gray      
}

\newcounter{JGHCommentsCounter}

\newcommand\anolrevised[1] {{ #1 }}

\newcommand{\indep}{\perp \!\!\! \perp}

\let\given\givenbase

\def\BibTeX{{\rm B\kern-.05em{\sc i\kern-.025em b}\kern-.08em
    T\kern-.1667em\lower.7ex\hbox{E}\kern-.125emX}}
\begin{document}

\title{Governing the Commons: Code Ownership and Code-Clones in Large-Scale Software Development}

\author*[1,2]{\fnm{Anders} \sur{Sundelin}}\email{anders.sundelin@bth.se}

\author[1]{\fnm{Javier} \sur{Gonzalez-Huerta}}\email{jgh@bth.se}
\equalcont{These authors contributed equally to this work.}

\author[3,4]{\fnm{Richard} \sur{Torkar}}\email{torkarr@chalmers.se}
\equalcont{These authors contributed equally to this work.}

\author[1]{\fnm{Krzysztof} \sur{Wnuk}}\email{krw@bth.se}
\equalcont{These authors contributed equally to this work.}

\affil*[1]{\orgdiv{Department of Software Engineering}, \orgname{Blekinge University of Technology}, \orgaddress{\postcode{371~79}~\city{Karlskrona}, \country{Sweden}}}

\affil[2]{\orgname{Ericsson Mobile Financial Services AB}, \orgaddress{\postcode{371~23}~\city{Karlskrona}, \country{Sweden}}}

\affil[3]{\orgdiv{Department of Computer Science and Engineering}, \orgname{Chalmers University of Technology and University of Gothenburg}, \orgaddress{\postcode{412~96}~\city{G{\"{o}}teborg}, 
\country{Sweden}}}

\affil[4]{\orgdiv{Stellenbosch Institute for Advanced Study (STIAS)}, \orgaddress{\city{Stellenbosch}, 
\country{South Africa}}}

\abstract{  \textbf{Context:} In software development organizations employing weak or collective ownership, different teams are allowed and expected to autonomously perform changes in various components.
This creates diversity both in the knowledge of, and in the responsibility for, individual components. 

  \textbf{Objective:} Our objective is to understand how and why different teams introduce technical debt in the form of code clones as they change different components. 

  \textbf{Method:} We collected data about change size and clone introductions made by ten teams in eight components which was part of a large industrial software system.
  We then designed a Multi-Level Generalized Linear Model (MLGLM), to illustrate the teams' differing behavior.
  Finally, we discussed the results with three development teams, plus line manager and the architect team, evaluating whether the model inferences aligned with what they expected.
  Responses were recorded and thematically coded.
  
  \textbf{Results:} The results show that teams do behave differently in different components, and the feedback from the teams indicates that this method of illustrating team behavior can be useful as a complement to traditional summary statistics of ownership.

  \textbf{Conclusions:} We find that our model-based approach produces useful visualizations of team introductions of code clones as they change different components.
  Practitioners stated that the visualizations gave them insights that were useful, and by comparing with an average team, inter-team comparisons can be avoided.
  Thus, this has the potential to be a useful feedback tool for teams in software development organizations that employ weak or collective ownership.
}

\maketitle

\section{Introduction}
\anolrevised{
Code clone detection and management has a long history in Software Engineering research~\citep{koschke2007survey,roy2007survey,RATTAN20131165,pate2013,Sobrinho2021}.
Though copy-pasted code initially functions like the original, issues may emerge as the code evolves, with clones diverging as they change at different rates, particularly when developers are unaware of them. \citet{TornhillXRays} states that two code clones that often, but not always, change together can be the source of problems.
In a study of five industrial and open-source systems, \citet{juergens2009code}~found that~52\% of all clones were inconsistently changed, and~15\% caused faults in the application.
For this reason, code clones are seen as a prominent and quite common Technical Debt Item (TDI). 

Technical Debt~\citep{cunningham1992wycash} (TD) is a metaphor borrowed from economics to explain the long-term consequences of sub-optimal design decisions taken to speed up the development process.
Technical Debt is classified into different categories~\citep{alves2016techdebt}, and the term Technical Debt Item (TDI) refers to a single occurrence of technical debt in software artifacts~\citep{Avgeriou2016,kruchten2019managing}.
}

\anolrevised{
Software development organizations often divide large projects into components, managed by different teams, and can take varying approaches to code ownership, from strong, to weak, or collective ownership~\citep{fowler2006code}.
In weak or collective ownership, different teams are allowed and encouraged to contribute to the same components.
}
\citet{ribeiro2016advantages} have identified advantages of weak ownership, such as 
\begin{enumerate*}[label=(\roman*)]
    \item knowledge distribution;
    \item increased backup pool of developers;
    \item lower rework; and
    \item better code quality
\end{enumerate*},
as well as disadvantages such as
\begin{enumerate*}[label=(\roman*)]
    \item increased conflict;
    \item increased errors and failures;
    \item lower understanding of the code; and
    \item increased development time
\end{enumerate*}
when studying shared code ownership in industrial contexts.

\anolrevised{
Traditional code ownership involves teams or individuals being responsible for the quality and upkeep of software artifacts~\citep{nordberg2003}.
However, in practice, it is challenging to assign responsibilities to development teams, because multiple teams, with different ``ownership stakes,'' frequently contribute to a single component.
}
Having a handful of contributors might cause friction in the development process by introducing overhead or degrading the quality of the product or its source code~\citep{avgeriou2016friction}.
\anolrevised{
For example, teams working in a component where they are not the main contributors might introduce more Technical Debt Items by being less careful with the code they produce, or they might introduce bugs due to their being unaware of the full consequences of their code changes. 
}
Therefore, the alignment between team expertise and tasks plays a critical role in the effectiveness and efficiency of the organization~\citep{newman2021building, baskarada2020microservices}.
\citet{sedano2016practice} emphasize that team code ownership is a feeling to be engendered, not a policy to be decreed.

In an article widely cited in economics and ecology, \citet{hardin1968tragedy} introduced the term ``The Tragedy of the Commons'', where he argued that common ownership in a land of finite resources will cause ruin and devastation for the whole population. Hardin argued in his essay for the centralization of the control of common-pool resources. 

However, \citet{dietz2003struggle} found five factors that support effective commons governance:
\begin{enumerate*}[label=(\roman*)]
    \item monitoring (including verification and understanding) of resources and human use of resources;
    \item moderate rate of change in resources, populations, technology, and social conditions;
    \item frequent face-to-face communication and dense social networks, which increase trust and allow people to experience the emotional reactions to distrust;
    \item the possibility of excluding outsiders at a relatively low cost; and
    \item the users themselves support effective monitoring and rule enforcement (i.e., the users understand the purpose of the rules).
\end{enumerate*}
We hypothesize that these factors apply to software code commons and that some of these factors can impact the mitigation of the ``code tragedy'', such as the accumulation of Technical Debt. 

\anolrevised{
The goal of this paper is to develop and evaluate a model to proactively monitor how Technical Debt growth varies by contributor in a large-scale industrial system, and how this aligns with component ownership.
We have built a model based on code clone introductions by teams, because clones are a common and easy TDI concept to grasp for most developers, with easy-to-grasp consequences.
Contributions were grouped by team because, in the studied organization, teams worked independently and did not have strong ownership (e.g., no mandatory inter-team code reviews).
We conducted a case study to collect data and build a Bayesian model that shows how code clones are introduced in components.
}

We created a Multi-Level Generalized Linear Model (MLGLM), based on a zero-inflated negative Binomial likelihood.
The model is fit onto a dataset consisting of $31007$ file changes made during 35 months in 8 code repositories by, in total, 10 development teams, belonging to an organization transitioning from collective to weak code ownership as it grew.
\anolrevised{
Based on causal reasoning, the model estimates the expected number of introduced code clones for a given change by a given team in a given repository.
The estimate is then used to visualize how teams behave in different repositories, and how they react to predictors such as existing complexity of, the number of existing duplicates in the changed file, or the size of the change.
}

To validate the reliability and usefulness of the model, we presented the main findings of the study to four of the studied teams in five focus-group sessions.
All teams agree that the model predictions and associated visualizations present a useful view of how teams have behaved, and they presented plausible reasons for why some teams ``stood out'' in certain repositories.

The paper is structured as follows:
Section~2 describes the background and related work, including the aforementioned OCAM model.
Section~3 describes the research methodology, including the design of the causal and statistical models.
Section~4 describes the results, both quantitative and qualitative.
Section~5 discusses our findings, followed by threats to validity in Section~6 and conclusions in Section~7.

\section{Background and Related Work}

A key part of the Agile and XP programming principles~\citep{Beck2000, abrantes2011agile}, the principle of collective code ownership have also been espoused by software craftsmanship authors~\citep[notably][]{Seibel2009coders,martin2011clean}.

Software engineering usually maps out individual ownership, based on metrics such as:
\begin{enumerate*}[label=(\roman*)]
    \item commits in a component,
    \item files authored or changed, or
    \item lines authored or removed.
\end{enumerate*}
In a study at Microsoft, \citet{bird2011don} defined ownership metrics based on the number of commits and examined the effect of these on software quality.
They looked at two large software projects:  Windows Vista and Windows 7, and explored whether the number of low-expertise developers and the proportion of ownership for the top owner have a relationship with both pre-release faults and post-release failures.
They discovered that more minor contributions (a proxy for weak code ownership) result in more pre- and post-release failures.
Other researchers at Microsoft replicated the study, looking in more detail at an intermediate level of granularity that lies between binaries and source files: code directories~\cite{greiler2015code}.
They also broadened the metrics describing code ownership to individual ownership for files, directories, and organizational ownership for files and directories.
Their results also confirmed that code ownership correlates with code quality. 

Several researchers have looked into code authorship and ownership in open-source software projects.
\citet{AVELINO201914} analyzed code authorship in 119 open-source projects, including the Linux kernel, and concluded that:
\begin{enumerate*}[label=(\alph*)]
    \item only a small portion of developers (26\%) makes significant contributions to the code base---this ratio is almost constant during the Linux kernel evolution; 
    \item the number of files per author is highly skewed—--a small group of top authors (2\%) is responsible for hundreds of files, while most authors (75\%) are responsible for at most 10 files; 
    \item most authors in Linux ($\approx76\%$) are specialists, and the ratio between specialists and generalists tends to be constant;
    \item authors with a high number of co-authorship connections tend to work with authors with fewer connections.
\end{enumerate*}
Similarly, \citet{foucault2015usefulness} investigated the relationship between Bird's code ownership metrics and software quality for seven open-source projects.
They confirmed the existence of a relationship between code ownership and software quality, but the relative importance of Bird's ownership metrics in multiple linear regression models is low compared to metrics such as the number of lines of code, the number of modifications performed over the last release, or the number of developers of a module.

\citet{maruping2009role} examined the role of team collective ownership and coding standards by surveying 73 software project teams belonging to a large US software development firm.
The authors analyzed 56 responses, comprising 509 software developers, and discovered that collective ownership and coding standards play a role in improving software project technical quality.

\citet{farago2015code} investigated the impact of code ownership, defined as the geometric mean of the number of authors of files in a commit, on maintainability for four open-source projects and discovered that code erosion is higher for source files modified by several developers in the past, compared to the files with clear ownership.
They concluded that files changed by more authors are more error-prone than those developed by fewer developers.

\citet{orru2016ownership} investigated the relationship between code ownership and refactoring activities in the Apache Ant software system.
Using Bird's notion of ownership as a ratio of changes performed by a single developer versus the total number of changes to a file, they derive two new metrics: Subjective Ownership and Relational Ownership.
They concluded that refactoring activities were positively correlated with both Subjective and Relational Ownership, although the relation is weaker in the Relational case.

\citet{borg2023uowns} analyzed 40 proprietary software repositories to understand the relationship between ownership, code quality, and issue resolution time.
They discovered that in low-quality source code, marginal owners need 45\% more time for small changes and 93\% more time for large changes.
Marginal owners are particularly hampered when working with low-quality source code, which leads to productivity losses. 

On the qualitative research front, \citet{ribeiro2016advantages} conducted an interview study with 19 participants to investigate the advantages and disadvantages of using shared code.
They found six advantages and six disadvantages of using shared code ownership, including improved knowledge distribution, increased conflicts between the team members, and increased development time. 

The Ownership and Contribution Model (OCAM) was formulated by~\citet{zabardast2022ownership}, and ranks teams using seven metrics to determine the alignment between contribution and ownership for a particular component.
The number of applied metrics is flexible, as is the granularity of contributors and components.
The authors validate the model in a longitudinal industrial case study in the paper.
In a follow-up study, \citet{Zabardast2022WFH} found that the relationship between ownership and contribution alignment was identified by the focus group participants as one of the potential causes of faster accumulation of technical debt.

In another study at Microsoft, \citet{herzig2014testownership} found that test suites whose owners are distributed across organization subgroups with long communication paths are negatively correlated with quality.
They recommended reviewing test suites concerning their organizational composition and favored subgroups having clear ownership of test suites.

\anolrevised{
  In addition to the studies mentioned in the introduction, several recent authors also studied code clones detection and management.
  \citet{AinSLRCodeClones2019} conducted a systematic literature review (SLR) of code clone detection tools, identifying 13 distinct tools in 54 selected papers, as well as 13 proposed future tools.
  \citet{quradaa2024systematic} surveyed the literature about the usage of recurrent neural networks in code clone detection, and concluded that LSTM techniques are the most commonly used.
  \citet{kaur2023systematic} surveyed the use of machine learning in code clone detection, reporting that Decision Tree, Random Forest, Bayesian Network, and Na\"{i}ve Bayes are the most popular machine learning algorithms, and that Deep Learning can be used to detect semantic clones.
   \citet{rongrong2019method} used different classifiers to recommend clones for refactoring, and concluded that Decision Trees and Bayesian Networks achieved the highest accuracy in recommending clones for refactoring, but Decision Tree was more stable.
  \citet{ZakeriNasrabadi2023} conducted another SLR on clone detection, and identified 136 primary studies, referring to 80 distinct tools, of which almost half support Java, and more than a third support C and C++.

  We identified two literature reviews about code clone evolution.
  \citet{pate2013} conducted an SLR on code clone evolution and found wide variation in the ratio of consistent changes to clones---between 11\% and 74\% of clones were found to be changed consistently.
  They concluded that researchers need to study human behavior, together with available data, to understand the evolution of code clones.
  \citet{Zhong2022} surveyed the literature on code clone evolution and found few studies on the visualization of clone evolution.
}

\citet{yu2012clones} studied the locations of code clones in two versions of the Linux kernel.
They found that clones mostly occurred between files close in the hierarchy, and found that the number of clones grew with the size of the kernel.
Although there are valid cases for duplicating code---in particular, if it changes for different reasons~\citep[see][]{martin2018clean, kapser2008cloning}---we follow the taxonomy of~\citet{alves2016techdebt} and place code clones in the \textit{Code Technical Debt} category, as several studies, such as~\citet{yu2012clones} have found that the majority of code clones arise due to simple copy-paste behavior.

Compared to the related work, we take a different approach.
Although code may be authored and committed by individuals, in proprietary software development, they tend to be organized in teams (that often have significant authority and freedom), and we hypothesize that team culture might have an impact on how developers approach code quality. 
Therefore, we map these individuals to the team they belong to at the time of authoring (or committing).
These teams are then used as categories to build a regression model, which can be used to simulate and visualize how teams introduce code clones in different components.
When presented with the results, both architects and development teams stress that they think the model for clone introductions is useful to illustrate and explain their behavior. 

\anolrevised{
However, when modeling other organizations or software development networks, other grouping factors might be used---regardless of \textit{how} the categories are conceived, the Bayesian model will use them when deriving the linear regression. 
Different organizations may warrant other grouping levels, such as individual authors, organizational sub-units, or different geographical sites or countries.
}

\section{Research Methodology}

The research presented in this paper follows a two-staged process, where we start with designing a causal model of how teams introduce clones in components, depending on their level of knowledge and how much they care for these components.
We then proceeded by using our causal model to guide us in collecting empirical data, following a case study research method.
We used two different sources of data:
\begin{enumerate*}[label=\roman*)]
  \item quantitative, for which we use archival analysis as data collection method to mine the software repositories of the studied organization to train models that predict the introduction of code clones in the studied software repositories; and
  \item qualitative, for which we use focus group interviews as a data collection method and open coding to analyze the data to validate the reliability and usefulness of the chosen model.
\end{enumerate*}

The remaining of the section is structured as follows: Subsection~\ref {sec:BuildingModel} provides the details on how the model was iteratively built, while Subsection~\ref{sec:case_study} reports on the empirical validation of the model predictions.

\subsection{Model Design}\label{sec:BuildingModel}

\subsubsection{Research Questions}

The motivation for our study is to explore how collective ownership of source code impacts its quality, in particular, the introduction of code clones. 
This led us to formulate the following research questions:

\begin{description}
\item[\textbf{RQ1}]
  Can we build a generalized linear model with acceptable accuracy and reliability to visualize team behavior regarding the introduction of code clones in different components?
\item[\textbf{RQ2}]
  What predictors are most likely to affect the rate of clone introduction, and how do these vary between teams and components?

\item[\textbf{RQ3}]
  Are the model predictions and their associated visualizations perceived as useful for the studied organization?
\end{description}
\noindent
For RQ1, we chose to design a Multi-Level Generalized Linear (MLGLM) model, predicting the number of introduced code clones in a file being changed, given various predictors.
We chose to focus on this particular class of Technical Debt Item (TDI), as we wanted a model that could predict team behavior without biasing how different classes of TDI should be weighted in the overall technical debt score.
Code clones are readily detected by tools such as \textsf{SonarQube}~\citep{roy2009clones}, and are also directly connected to the clean--code principle \textit{DRY - Don't Repeat Yourself}~\citep{hunt2000pragmatic, martin2011clean}, that have been shown to be an important indicator of technical debt by experienced developers~\citep{Ljung2022}.

For RQ2, we designed several models and compared their out-of-sample prediction capabilities using the LOO-CV metric~\citep{vehtari2017practical}.

For RQ3, we conducted five focus group interviews with four of the studied teams---one meeting each with three development teams, plus opening and closing presentations with the architect team.
The meetings were recorded and transcribed; anonymized transcriptions are available upon request.
We used open coding to summarize the findings of the teams.
Based on team feedback, we also designed a follow-up model building of clone removal behavior.

\subsubsection{Measuring the Degree of Ownership Alignment}

OCAM was originally defined as a flexible model to assess the degree of alignment between \textit{formal ownership} and team contribution to a given repository~\citep{zabardast2022ownership}.
The OCAM model allows metrics and organizational granularity to be configured according to specific needs.
As the studied organization was large and some developers switched teams, we decided to keep the team granularity from the original OCAM model.

\begin{table}
    \caption{Used OCAM model metrics.}
    \label{tab:ocam_metrics}
    \begin{tabular}{r|l}
        \textbf{Metric} & \textbf{Description} \\
        \hline
        \textsf{N} & Number of commits \\
        \textsf{CHURN} & Sum of \texttt{max(added, removed)} LOC \\
        \textsf{ADD\_COMP}  & Sum of added McCabe complexity \\
        \textsf{REM\_COMP}  & Sum of removed McCabe complexity 
    \end{tabular}
\end{table}

Table~\ref{tab:ocam_metrics} contains descriptions of the OCAM metrics that we collected.
We kept the following original OCAM metrics:
\begin{enumerate*}[label=(\roman*)]
    \item the number of commits;
    \item code churn---defined as the sum of \textsf{max(added, removed)} lines for each changed file;
    \item added complexity (McCabe metric); 
    \item removed complexity.
\end{enumerate*}
We had to disregard the metrics regarding tickets and pull requests, since the studied organization did not keep track of these metrics for changes originating between teams.

We used the OCAM model in the second stage of our research, comparing and contrasting its results with the results from our model, as perceived by the development teams and architects. 

\subsubsection{Predicting Code Clone Introduction}

In the design of the statistical model for clone introduction, we follow the method outlined by~\citet{mcelreath2020statistical}, which starts with stating your assumptions and the domain knowledge in the form of a Directed Acyclic Graph~(DAG)~\citep{pearl2009causality, furiaTF23causal}.

\begin{figure}
  \centering
  \includegraphics[width=\columnwidth]{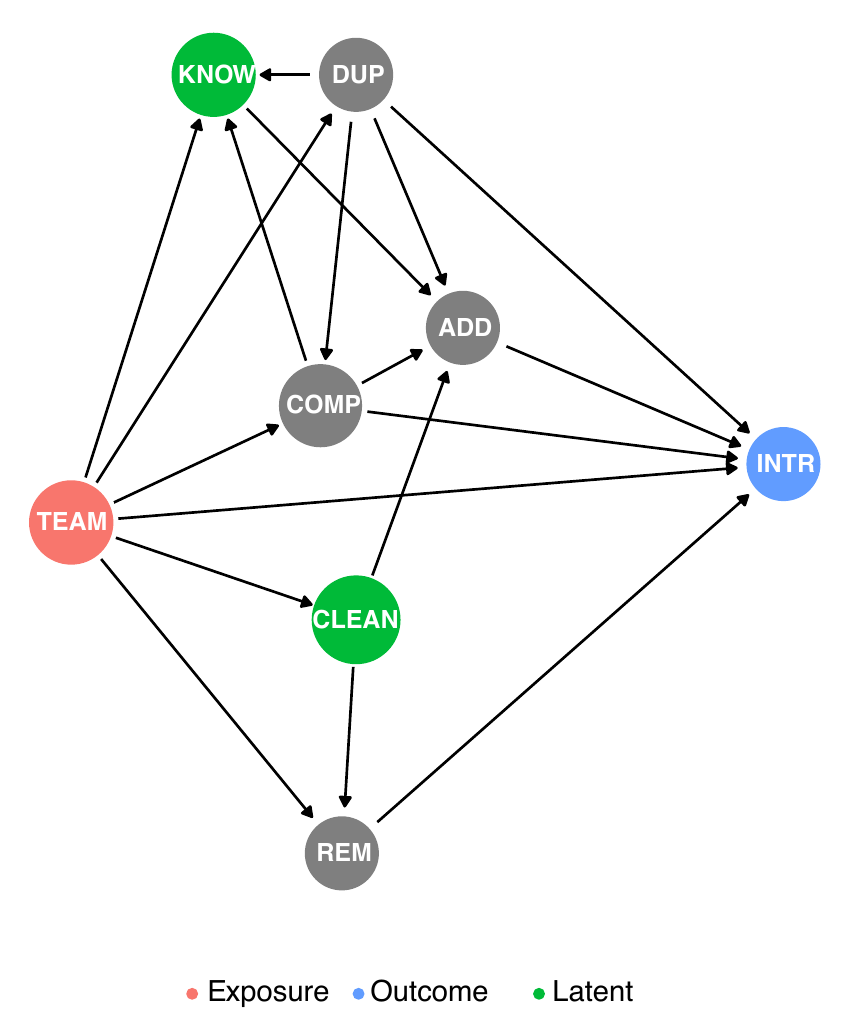}
  \caption{Assumed causal model of code clone introduction.}
  \label{fig_causal_dag}
\end{figure}
  
Figure~\ref{fig_causal_dag} shows a causal DAG containing the variables and their dependencies in our model of clone introductions.
Two variables are unknown (and therefore unmeasured) and are marked as latent---\texttt{CLEAN} and \texttt{KNOW}.
In this DAG,\footnote{DAG motivation is available at \url{https://docs.google.com/spreadsheets/d/1vV6utjEEgmELib9ihplflOjhjVLPDXvHWbUG49cb-qQ}} every change to each source code file is modeled according to the following variables:

\begin{description}
    \item[INTR] (outcome)
      measures how many duplicates were introduced by the change to the file.
      This is the outcome variable of the model.
      The lower bound is zero, and the upper limit is bounded by the size of the changed file (though the duplicate detection tool will consider several adjacent identical lines a single duplicate).
    
    \item[TEAM] (exposure)
      indicates the team that committed the change to the master branch, as a categorical predictor.
      Furthermore, we assume that behavior will vary between teams and that a single team will behave differently in different components (repositories).
      We map the committer to the team at the time of the commit, based on sampled organization charts from a team wiki page, which contains current and historical team compositions.
      
    \item[ADD] (adjustment set)
      measures the number of lines that were added to the file in this particular change.
      The lower bound is zero, and there is no theoretical upper bound (although in practice, the used compiler will impose some fixed limit).

    \item[REM] (adjustment set)
      measures the number of lines that were removed from the file in this particular change.
      The lower bound is zero, and the upper bound is limited by the size of the file before the change.
    
    \item[DUP] (adjustment set)
      measures the number of existing duplicates in the file before the change, as measured by \textsf{SonarQube}.
      The lower bound is zero, and the theoretical upper bound is limited by the size of the changed file (based on how the tool calculates duplicates).

    \item[COMP] (adjustment set)
      measures the complexity of the file according to~\citet{mccabe1976complexity}, as measured by \textsf{SonarQube}, after the change.
      The lower bound is zero, and there is no theoretical upper bound.

    \item[\texttt{KNOW}]
      is a latent (unmeasured and unknown) metric, representing the knowledge the team committing the change has in the file, and in the component to which the changed file belongs.
      As complex files---possibly with many existing duplicates---are likely to be harder to comprehend, the prior knowledge is affected by the DUP and COMP metrics.
      Knowledge affects the number of added lines, as we posit that developers make more copy-paste changes in components where they lack domain knowledge.

    \item[\texttt{CLEAN}]
    is a latent (unmeasured and unknown) metric, representing the tendency of the team committing the change to clean up the code they touch or come across when they make changes.
    A team being formally responsible for the quality of---or with a high ownership stake in---a given component would be expected to clean up more (e.g., via refactorings) than a sporadically contributing team.

\end{description}

\noindent
The main causal effect we want to study is how team knowledge and cleanliness affects the number of duplicates introduced for a given number of added lines.
As we cannot measure these metrics objectively, we instead measure how the team and repository combination affect the influence that the other predictors (i.e., the number of duplicates in the file, the complexity, and the number of lines added or removed) have on the amount of duplicates being introduced.

The specified DAG implies the following conditional independencies:
$\mathrm{REM} \indep \mathrm{COMP} \given \mathrm{TEAM}$ and $\mathrm{REM} \indep \mathrm{DUP} \given \mathrm{TEAM}$ ($\indep$ is the symbol for independence, and $\given$ is the symbol for `given').

These independencies can be tested from the collected data, and while the number of removed lines does seem to be independent of the existing complexity, there might be a weak association between $\mathrm{REM}$ and $\mathrm{DUP}$ ($\mu=0.13$, 95\%~$\mathrm{CI}=[0.11;0.14])$.
This might indicate that (some) teams are more likely to remove lines in files with many duplicates.
Thus, we do not rule out a causal path between $\mathrm{DUP}$ and $\mathrm{REM}$.
However, the existence of this path would not change our model or conclusions since, most likely, this path might only influence how the duplicates are removed, i.e., a file with many duplicates might receive changes that mainly focus on removing lines of code, or at least modifying them, aiming at removing duplicates.

\subsubsection{Data Collection}

\begin{algorithm}
\DontPrintSemicolon
\SetAlgoLined
\SetKwFunction{AUTHOR}{AuthoredBy}
\SetKwFunction{COMMITTER}{CommittedBy}
\SetKwFunction{ATIME}{AuthoredDate}
\SetKwFunction{CTIME}{CommittedDate}
\SetKwFunction{TEAMATDATE}{FindTeam}
\SetKwFunction{FILES}{Files}
\SetKwFunction{ADDED}{Added}
\SetKwFunction{REMOVED}{Removed}
\SetKwFunction{COMPLEX}{Complexity}
\SetKwFunction{DUPL}{Duplicates}
\SetKwData{filestate}{filestate}
\SetKwData{T}{d}
\SetKwData{AUTH}{author}
\SetKwData{COMM}{committer}
\SetKwData{TEAM}{team}
\SetKwData{ADD}{add}
\SetKwData{REM}{rem}
\SetKwData{COMP}{comp}
\SetKwData{DUP}{dup}
\SetKwData{INTROD}{introd}
\KwData{$C$ commits$: C_i$ precedes $C_{i+1}$}
\KwResult{List of data for changed files}
 \filestate $\gets ()$\;
 \For{$C_i \in C$}{
   $\AUTH \gets $ \AUTHOR{$C_i$}\;
   $\T_{auth} \gets $ \ATIME{$C_i$}\;
   $\COMM \gets $ \COMMITTER{$C_i$}\;
   $\T_{comm} \gets $ \CTIME{$C_i$}\;
   $\TEAM_{a} \gets $ \TEAMATDATE{$\AUTH, \T_{auth}$}\;
   $\TEAM_{c} \gets $ \TEAMATDATE{$\COMM, \T_{comm}$}\;
   \For{$F_{j,i} \in$ \FILES{$C_i$}}{
     $\ADD \gets$ \ADDED{$F_{j,i}$}\;
     $\REM \gets$ \REMOVED{$F_{j,i}$}\;
     $\COMP \gets$ \COMPLEX{$F_{j,i}$}\;
     $\DUP_{prev} \gets$ \DUPL{$F_{j,i-i}$}\;
     $\DUP_{curr} \gets$ \DUPL{$F_{j,i}$}\;
     $\Delta \gets \DUP_{curr} - \DUP_{prev}$\;
     \lIf{$\Delta > 0$}{$\INTROD \gets ~\Delta$}
     \lElse{$\INTROD \gets ~0$}
     $\filestate \gets \filestate + (F_{j,i}, \TEAM_a, \TEAM_c, \\
     \ADD, \REM, \COMP, \DUP_{prev}, \INTROD)$\;
     
  }
}
\Return{\filestate}
\caption{Calculating introduced clones for changed files in a repository.}
\label{alg:file_change}
\end{algorithm}

We used Algorithm~\ref{alg:file_change} to calculate the required metrics for each changed file, in each commit, per repository. 
List $C$ contains each commit, in order of application to the master branch for each studied repository.
Commit data is retrieved via the functions \texttt{AuthoredBy($C$)}, \texttt{AuthoredDate($C$)}, \texttt{CommittedBy($C$)} and \texttt{CommittedDate($C$)}. 
In the \textsf{git} version control system, the \texttt{author} is normally the person initiating the change, and the \texttt{committer} is the person triggering the commit which merges the change to the main branch.

We used historical organization charts to determine the developer--team affiliation at a given time. This information is returned by the \texttt{FindTeam(A, D)} function.
\texttt{Files($C$)} are the files changed in commit $C$.
\texttt{Added($F_{j,i}$)} and \texttt{Removed($F_{j,i}$)} are the number of added and removed lines of file $F_j$, in commit $C_i$, as counted by the \textsf{git} command.
\texttt{Complex($F$)} and \texttt{Duplicates($F$)} are the McCabe complexity and number of duplicated code blocks, as measured by the \textsf{SonarQube} tool.
We use the $F_{j,i-1}$ notation to denote the file state in commit $C_{i-1}$, regardless of whether the file changed in that commit or not.
The number of introduced duplicated code blocks is given by~$\Delta$, in case it is positive.
The complete anonymized data set is available in the replication package~\citep{epkanol_2024_11357296}.

\subsubsection{Simulation and Initial Model Design}

Based on our prior domain knowledge, and supported by empirical findings such as~\citet{Zabardast2020refactoring}, who found that most file changes neither introduced nor removed Technical Debt Items, we assumed that most file changes would not introduce any additional duplicates. 
As we were modeling a count ($\geq 0$) of added duplicates, our initial model used the Zero-Inflated Poisson distribution, which is the maximum entropy distribution for counting independent events, having a known expected value.

However, after simulating and validating with collected data from the first repository, we found the constraints of the Poisson distribution, i.e., equal variance and expected value ($\sigma^2 = \lambda$), to be unsuitable for our domain.
The conventional next choice of model is the Negative-Binomial distribution, which is parameterized via two parameters: $\mu$, the expected value (mean), and $\phi$, the shape parameter that adjusts the variance via the formula $\sigma^2 = \mu + \frac{\mu^2}{\phi}$.
As recommended by~\citet{mcelreath2020statistical}, we used a logarithmic link function, $\mathrm{log}(\mu) = \beta_0 + \sum_i{\beta_i P_i}$ to tie our linear predictors to the parameters of the Negative-Binomial.
The zero-inflation part is realized via a Bernoulli trial, with the probability of `success' (zero-inflation active) tied to the linear predictor via a logit (log-odds) link function, $\mathrm{logit}(p) = \mathrm{log}(\frac{p}{1-p})$.

Following the recommendations by~\citet{mcelreath2020statistical} and Gelman,\footnote{\url{https://statmodeling.stat.columbia.edu/2019/08/21/you-should-usually-log-transform-your-positive-data/}} we transformed our data to fit within the useful range of the link functions.
In theory, the logit function is unbounded, but in practice, the useful values lie within $\pm 10$, as $\mathrm{logit}(4.5~\cdot~10^{-5})~\approx~-10$ and $\mathrm{logit}(0.99995)~\approx~10$.
A similar argument can be made for the log link function, which causes the mean value of the Negative-Binomial likelihood to grow with the exponent of the linear predictors.

Based on our causal assumptions (Fig.~\ref{fig_causal_dag}), besides the team and repository categorical predictors, we considered adding four numerical predictors: added lines (ADD), removed lines (REM), existing duplicates (DUP) and existing complexity (COMP).
Following the DAG, all numerical predictors were bounded by zero, without any obvious upper bound.
Furthermore, exploratory analysis of collected data showed that all four predictors were right-skewed,\footnote{Most changes add a few lines of code, remove a few lines of code, and the complexity of the files is in general low, with fewer data points with high levels for any of these variables.} which caused us to base the linear model based on the logarithm of the predictor value plus one (as $\mathrm{log}(0+1)~=~0$).
We scaled and centered the resulting logarithm by subtracting the mean and dividing by the standard deviation of the logarithm, to aid prior selection and improve model fitting, as suggested in ~\citep{mcelreath2020statistical}.
This means that a change of one unit in the scaled predictor corresponds to a change in the magnitude of the unscaled predictor equal to the observed standard deviation of the magnitude.

This leaves us with four potential numerical predictors:

\begin{description}
    \item{\textbf{A}} is the scaled and centered natural logarithm of the number of added lines (ADD).
    \item{\textbf{R}} is the scaled and centered natural logarithm of the number of removed lines (REM).
    \item{\textbf{C}} is the scaled and centered natural logarithm of the existing McCabe complexity of the changed file, as calculated by \textsf{SonarQube} (COMPLEX).
    \item{\textbf{D}} is the scaled and centered natural logarithm of the existing number of duplicates in a changed file, as calculated by \textsf{SonarQube} (DUP).   
\end{description}
\noindent
Depending on the number of predictors we include, we will get models of varying complexity.

A crucial step in model design is conducting prior predictive checks~\citep{mcelreath2020statistical}.
For Bayesian inference to work efficiently, the selected priors should be wide enough to allow for reasonable outcomes, but not too wide, as this would force the model to explore outcomes not consistent with domain-specific knowledge.
In our context, we know, based on the design of the duplicate detection tool,\footnote{SonarQube, \url{https://www.sonarsource.com/products/sonarqube}} that the number of introduced duplicates must be less than the number of existing lines in a file.
Thus, regardless of the parameter values fed to the model, we would find outcomes ranging in the tens of thousands to be implausible.

Both domain knowledge simulation and prior predictive checks were conducted before actual data collection, and both are available in the replication package~\citep{epkanol_2024_11357296}.

\subsubsection{Model Design and Quality Assessment}

We built three models of increasing complexity to compare and assess their results.
We used the \textsf{R} programming language\footnote{\url{https://www.r-project.org/}} and the framework \textsf{BRMS}~\citep{burkner2017brms, burkner2018brms} as a front-end for the Hamiltonian Monte Carlo (HMC) framework provided by \textsf{Stan}~\citep{carpenter2017stan}.

\subsubsection{Model Structure}

All three models use a Zero-Inflated Negative-Binomial distribution, and differ in complexity only in the number of parameters used in the linear predictors.

\begin{equation}
    p(y | \mu, \phi, \xi) = \begin{cases} 
      \xi + (1-\xi)\cdot \text{NB}(0 | \mu, \phi) & \text{if } y = 0\\ 
      (1-\xi)\cdot\text{NB}(y | \mu, \phi) & \text{if } y \neq 0
    \end{cases}
    \label{eqn_zinb}
\end{equation}
\noindent
A Zero-Inflated Negative-Binomial model is defined by Equation~\ref{eqn_zinb}, where NB is the Negative-Binomial distribution with mean (i.e., location) parameter $\mu$ and a shape parameter $\phi$, which adjusts the variance via the formula~$\sigma^2 = \mu + \frac{\mu^2}{\phi}$.
The parameters $y$, $\mu$, $\phi$, and $\xi$ can either be modeled as population-level parameters (i.e., the same distribution for all values in the input data set), or may depend on observation~$i$.

We will adopt the notation favored by \citet{mcelreath2020statistical}, where an~$i$ suffix indicates that the parameter is associated with observation~$i$ from the observed data.
That is: $\mu_i$ means the $\mu$ value for observation~$i$; $\beta_{0,\tau[i]}$ is the intercept parameter offset associated with the team in observation~$i$; $\sigma_{\tau}$~means the standard deviation for team~$\tau$; and~$\sigma_{\tau:\chi}$ is the standard deviation associated with team~$\tau$ in component (repository)~$\chi$.

In all our models, we leave the shape parameter, $\phi$, independent of observations.

\subsubsection{Intercept-Only Model $\mathcal{M}_0$}

The first, simplest model $\mathcal{M}_0$, ignores all predictors except the team and repository categories.
This is a common choice as a baseline model, for comparison with models of higher complexity.
It is specified as:

$\mathcal{M}_0$
\begin{equation}
  \begin{aligned}
    \log(\mu_i) &= \beta_{0,i} \\
    \text{logit}(\xi_i) &= \gamma_{0,i}
  \end{aligned}
  \label{eqn_zinb_m0_mu_xi}
\end{equation}

where 

\begin{equation}
  \begin{aligned}
    \beta_{0,i} &= \beta_{0} + \beta_{0,\tau[i]} + \beta_{0,\tau:\chi[i]} \\
    \gamma_{0,i} &= \gamma_{0} + \gamma_{0,\tau[i]} + \gamma_{0,\tau:\chi[i]}      
  \end{aligned}
  \label{eqn_zinb_m0_team}
\end{equation}

with priors

\begin{equation}
    \begin{aligned} 
    \beta_{0} &\sim \text{Normal}(0,0.5) \\
    \sigma_{\tau} &\sim \text{Weibull}(2, 0.25) \\
    \sigma_{\tau:\chi} &\sim \text{Weibull}(2, 0.25) \\
    \phi &\sim \text{Gamma}(0.5,0.1)
    \end{aligned}
  \label{eqn_zinb_m0_priors}
\end{equation}
\noindent
Model $\mathcal{M}_0$ assumes that the logarithm of the mean ($\mu_{i}$) of the Negative-Binomial for observation $i$ is composed of one population-level parameter $\beta_{0}$, plus two adjustments to this value; $\beta_{0,\tau[i]}$, adjusting the intercept per team (regardless of the repository), and $\beta_{0,\tau:\chi[i]}$, additionally adjusting the intercept depending on the team--repository combination.

The same structure is used for the probability ($\xi_i$) of seeing inflated zeros; in this case, $\gamma_{0}$ represents the population-level intercept, adjusted by $\gamma_{0,\tau[i]}$ and $\gamma_{0,\tau:\chi[i]}$.
The model assumes that the shape parameter is independent of observations.

\subsubsection{A More Complex Model $\mathcal{M}_1$}

Following our DAG (Fig.~\ref{fig_causal_dag}), we posit that the causal model of introduced clones would at least include the added and removed lines predictors, whose scaled magnitude is represented by A and R, with team- and repository-level variation only on the intercept and the removed lines (R) predictor.

Our second model then becomes:

$\mathcal{M}_1$
\begin{equation}
  \begin{aligned} 
    \log(\mu_i) &= \beta_{0,i} + \beta_A A_i + \beta_{R,i} R_i \\
    \text{logit}(\xi_i) &= \gamma_{0,i} + \gamma_A A_i + \gamma_{R,i} R_i
  \end{aligned}
\end{equation}

Here, the intercept ($\beta_{0,i}$) and the slope for the R predictor ($\beta_{R,i}$) are composed of three components, incorporating the team- and team-per-repository-level differences:

\begin{equation}
  \begin{aligned}      
    \beta_{0,i} &= \beta_{0} + \beta_{0,\tau[i]} + \beta_{0,\tau:\chi[i]} \\
    \beta_{R,i} &= \beta_{R} + \beta_{R,\tau[i]} + \beta_{R,\tau:\chi[i]} 
  \end{aligned}
\end{equation}

In contrast, the slope $\beta_A$ of the added lines predictor is assumed to be constant between teams and repositories, meaning that as the number of added lines grow, the rate of introducing duplicates increases by the same amount for all teams and repositories, if the other predictors are kept constant.
However, as the model allows the intercepts to vary between teams and repositories, each team might still have a different base rate of clone introduction in each repository.

As we let both the slope and the intercepts vary, we need to incorporate prior information for the correlation between these parameters.
The offsets per team ($\beta_{R,\tau[i]}$) and team--repository ($\beta_{R,\tau:\chi[i]}$) are modeled as multi-variate-normal (MVN) models, with a zero mean---because they are offsets---and standard deviation matrices ($\Sigma_{R,\tau}$, $\Sigma_{R,\tau:\chi}$) decomposed by diagonal matrices and correlation matrices ($\Omega_{R,\tau}$, $\Omega_{R,\tau:\chi}$).
We use the recommended Lewandowski-Kurowicka-Joe distribution for the correlation matrices; our choice of the $\mathrm{LKJ}(2)$ prior ensures that our model is mildly skeptical of extreme correlations between intercepts and slopes.

This means that our priors are defined as:

\begin{equation}
  \label{eq_full_priors}
  \begin{aligned} 
    \beta_{0}       &\sim \text{Normal}(0,0.5) \\
    \beta_{A}       &\sim \text{Normal}(0,0.25) \\
    \beta_{R}       &\sim \text{Normal}(0,0.25) \\
    \beta_{R,\tau[i]} &\sim \text{MVN}(0,\Sigma_{R,\tau}) \\
    \Sigma_{R,\tau}  &\sim \text{diag}(\sigma_{\tau})\Omega_{\tau}\text{diag}(\sigma_{\tau}) \\
    \sigma_{\tau}    &\sim \text{Weibull}(2, 0.25) \\
    \Omega_{\tau}    &\sim \text{LKJ}(2) \\ 
    \beta_{R,\tau:\chi[i]} &\sim \text{MVN}(0,\Sigma_{R,\tau:\chi}) \\
    \Sigma_{R,\tau:\chi}  &\sim \text{diag}(\sigma_{\tau:\chi})\Omega_{\tau:\chi}\text{diag}(\sigma_{\tau:\chi}) \\
    \sigma_{\tau:\chi}    &\sim \text{Weibull}(2, 0.25) \\
    \Omega_{\tau:\chi}    &\sim \text{LKJ}(2) \\
    \phi               &\sim \text{Gamma}(0.5, 0.1)
  \end{aligned}
\end{equation}

The zero-inflation predictors ($\gamma$) use identical structures and priors:

\begin{equation}
  \begin{aligned}
    \gamma_{0,i} &= \gamma_{0} + \gamma_{0,\tau[i]} + \gamma_{0,\tau:\chi[i]} \\
    \gamma_{R,i} &= \gamma_{R} + \gamma_{R,\tau[i]} + \gamma_{R,\tau:\chi[i]}
  \end{aligned}
\end{equation}

\subsubsection{Full Causal Model $\mathcal{M}_2$}

Following the complete DAG in Fig.~\ref{fig_causal_dag}, we posit that the existing complexity and number of duplicates in a file impact the number of duplicates that a particular team introduces, especially if the team is unfamiliar with the component where the file resides.

The fact that the existing number of duplicates (DUP, part of code technical debt) in a file will impact the number of duplicates that developers introduce while changing the same file is consistent with the \textit{Broken Window Theory} phenomenon, first described in a Software Engineering context by \citet{hunt2000pragmatic} and validated by \citet{leven2022}.

With this reasoning in mind, we decided to also incorporate the existing complexity and number of duplicates into our model:

$\mathcal{M}_2$
\begin{equation}
  \begin{aligned}
    \log(\mu_i) &= \beta_{0,i} + \beta_A A_i + \sum_{P \in \{R,C,D\}} \beta_{P,i} P_i \\
    \text{logit}(\xi_i) &= \gamma_{0,i} + \gamma_A A_i + \sum_{P \in \{R,C,D\}} \gamma_{P,i} P_i
  \end{aligned}   
  \label{eqn_zinb_m2}
\end{equation}

Compared with $\mathcal{M}_1$, our model now has two additional predictors ($C$ and $D$), representing the scaled magnitude of the complexity (COMPLEX) and existing duplicates (DUP) in the file.
The $\beta_{C,i}$ and $\beta_{D,i}$ coefficients, just like $\beta_{0,i}$ and $\beta_{R,i}$ in model $\mathcal{M}_1$, are composed of three components, as is $\gamma_{C,i}$ and $\gamma_{D,i}$.

\begin{equation}
  \begin{aligned}      
    \beta_{0,i} &= \beta_{0} + \beta_{0,\tau[i]} + \beta_{0,\tau:\chi[i]} \\
    \beta_{R,i} &= \beta_{R} + \beta_{R,\tau[i]} + \beta_{R,\tau:\chi[i]} \\
    \beta_{C,i} &= \beta_{C} + \beta_{C,\tau[i]} + \beta_{C,\tau:\chi[i]} \\
    \beta_{D,i} &= \beta_{D} + \beta_{D,\tau[i]} + \beta_{D,\tau:\chi[i]} \\
    \gamma_{0,i} &= \gamma_{0} + \gamma_{0,\tau[i]} + \gamma_{0,\tau:\chi[i]} \\
    \gamma_{R,i} &= \gamma_{R} + \gamma_{R,\tau[i]} + \gamma_{R,\tau:\chi[i]} \\
    \gamma_{C,i} &= \gamma_{C} + \gamma_{C,\tau[i]} + \gamma_{C,\tau:\chi[i]} \\
    \gamma_{D,i} &= \gamma_{D} + \gamma_{D,\tau[i]} + \gamma_{D,\tau:\chi[i]} 
  \end{aligned}
  \label{eqn_zinb_m2_beta}
\end{equation}

We used the same priors as in model $\mathcal{M}_1$, but also included $C$ and $D$: 
\begin{equation}
    \begin{aligned} 
    \beta_{0} &\sim \text{Normal}(0,0.5) \\
    \forall P \in \{A, R, C, D\}: \beta_{P} &\sim \text{Normal}(0,0.25) \\
    \forall P \in \{R, C, D\}: \beta_{P,\tau[i]} &\sim \text{MVN}(0,\Sigma_{P,\tau}) \\
    \forall P \in \{R, C, D\}: \Sigma_{P,\tau} &\sim \text{diag}(\sigma_{\tau})\Omega_{\tau}\text{diag}(\sigma_{\tau}) \\
    \sigma_{\tau} &\sim \text{Weibull}(2, 0.25) \\
    \Omega_{\tau} &\sim \text{LKJ}(2) \\ 
    \forall P \in \{R, C, D\}: \beta_{P,\tau:\chi[i]} &\sim \text{MVN}(0,\Sigma_{P,\tau:\chi}) \\
    \forall P \in \{R, C, D\}: \Sigma_{P,\tau:\chi} &\sim \text{diag}(\sigma_{\tau:\chi})\Omega_{\tau:\chi}\text{diag}(\sigma_{\tau:\chi}) \\
    \sigma_{\tau:\chi} &\sim \text{Weibull}(2, 0.25) \\
    \Omega_{\tau:\chi} &\sim \text{LKJ}(2) \\ \phi &\sim \text{Gamma}(0.5, 0.1)
    \end{aligned}
  \label{eqn_zinb_m2_priors}
\end{equation}

As for model $\mathcal{M}_1$, we used identical structure and priors for the predictors for the zero-inflation component ($\gamma$).

\begin{figure}
\centering
\includegraphics[width=0.95\linewidth]{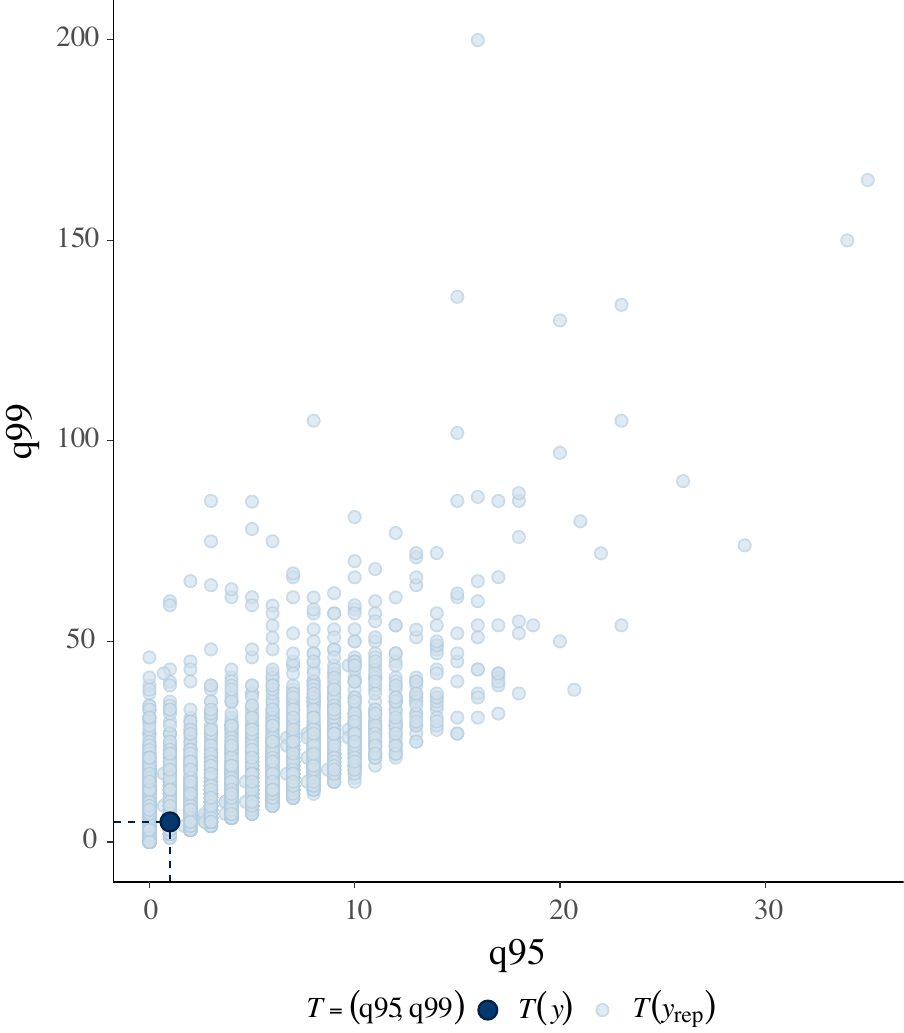}
\caption{Prior predictive check plot of 95th vs. 99th percentile of introduced code clones. $T(y_{rep})$:~Prior predictions. $T(y)$:~Observations ($Q_{95}~=~1$, $Q_{99}~=~5$). Prior predictions are consistent with observations, and not unreasonable.}
\label{fig_prior_pc_q95_vs_q99}
\end{figure}

We used visualizations such as the plot in Fig.~\ref{fig_prior_pc_q95_vs_q99} to aid prior selection for all our models.
\citet{gelman2020bayesian} recommends avoiding bounded priors such as the uniform distribution, and instead use priors consistent with domain knowledge.
In Fig.~\ref{fig_prior_pc_q95_vs_q99}, the $x$-axis depicts the $95$th percentile and the $y$-axis depicts the $99$th percentile of predictions made using only prior information, with predictor values used from our data.
Based on the figure, we conclude that using these priors, the model would expect $95$\% of the observations to range between $0$ and about $20$ introduced duplicates (the observed count is $1$, indicated by $T(y)$ in the figure).
Likewise, the model would expect $99$\% of the observations to range between $0$ and about $80$ introduced duplicates, with the more likely outcomes being between $0$--$40$.

We used the following criteria to select appropriate priors:
\begin{itemize}
    \item Based on empirical findings~\citep{Zabardast2020refactoring} and experience, we would find it implausible that the majority of file changes would result in new duplicates. Our priors put the highest probability at about $80$\% zeros, but expect anything from $40$\% to $100$\% zeros (with smaller probabilities at the extremes).
    \item The maximum estimated value (i.e., the number of introduced clones the model would expect, based only on prior information) for these priors range from $\approx100$ to $\approx20000$ duplicates, with higher probabilities at the lower end of the scale, and very little probability over $1000$.
    \item As shown in Fig.~\ref{fig_prior_pc_q95_vs_q99}, the 95th percentile is expected to range from $0$ to $\approx20$ and the 99th percentile is expected to range from about $0$ to $\approx80$, though more extreme values are also possible, albeit less likely.
\end{itemize}

The replication package contains more prior predictive plots, including predictions on group-level data, such as teams and repositories.

Based on our exploratory data analysis, we found that our data contained an excess amount of zeros (around 93\%), so traditional measures of central tendency (mean and median) would not be of much use.
We opted instead to use Q95 and Q99 metric---that is the 95th and 99th percentile metrics. 

After fitting the model, and performing standard sampling checks,\footnote{See the replication package for details.} we assessed the model fit by approximating leave-one-out cross-validation~(LOO-CV) using Pareto-smoothed importance sampling (PSIS), as recommended by~\citet{vehtari2017practical}.
Instead of doing full leave-one-out cross-validation, which would require refitting the model once for each data point, the LOO-CV algorithm calculates the leave-one-out posterior distribution using Pareto-smoothed importance sampling (PSIS) by approximating the importance of each data point on the final posterior distribution.
The result is reported for each data point as the Pareto $k$-value.
Data points with Pareto $k$-values higher than 0.7 are considered outliers and should be investigated further.
In our case, model $\mathcal{M}_2$ contained 12 data points with Pareto $k$ values exceeding 0.7.
Using the \textsf{reloo} function,\footnote{\url{https://discourse.mc-stan.org/t/clarification-about-the-purpose-of-reloo-in-the-loo-function/8742}} the model was refit once per suspicious data point, concluding that the real Pareto $k$ values for all data points were below the recommended threshold.
This completed our diagnostics of the models.

\begin{figure}
\centering
\includegraphics[width=0.95\linewidth]{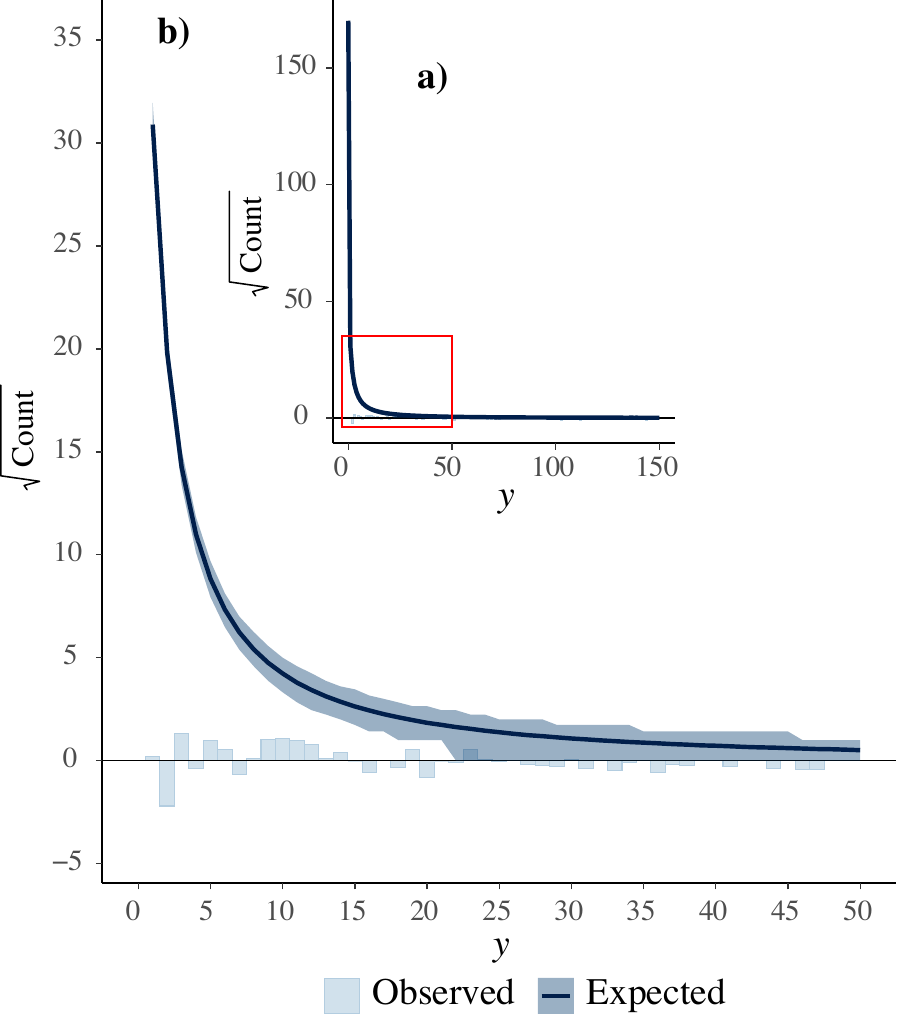}
\caption{Suspended rootogram for model $\mathcal{M}_2$. \\
Subplot a): expected count 0-150 \\
Subplot b): expected count 1-50; enlargement of the red square}
\label{fig_rootogram}
\end{figure}

However, a well-diagnosed model might still produce bad inferences, in case it does not fit the data well.
The conventional tool to assess model fit is to plot and quantify residuals (the difference between predicted and observed values) but for zero-inflated models like ours, the recommended way is instead to use the rootogram method~\citep{kleiber2016visualizing}.
In a rootogram, the $y$-axis contains the square root of the expected frequency and the $x$-axis contains the predicted outcome count.
In a suspended rootogram, the difference between the expected and actual frequency is shown as the light-blue \textit{Observed} histogram, while the \textit{Expected} line (and surrounding credible interval) contains the expected frequency.

Figure~\ref{fig_rootogram} contains two subplots of rootograms in the suspended style, for model $\mathcal{M}_2$.
Subplot \textbf{a)} contains predictions up to $y = 150$, and all of the predicted zeros (as $\sqrt{30000} \approx 173$). 
At this scale, it is hard to draw any conclusions.
Therefore, we created subplot~\textbf{b)}, which zooms in on predictions between 1 and 50.
At this scale, we see that the model slightly underestimates introductions of 2 duplicates, and slightly less overestimates 3 duplicates.
The rest of the expected counts fit progressively better.
Overall, the model seems to fit the data well---this is also indicated by the narrow prediction interval around the expected frequency.

\subsubsection{Model Comparison}

The LOO-CV algorithm can also be used to compare and rank the models, as shown in Table~\ref{tab:loo_compare}.
This indicates that model $\mathcal{M}_2$ outperforms the simpler model $\mathcal{M}_1$ by about eight standard deviations ($\frac{205.9}{24.7}$), which is a significant amount.
The intercept-only model, $\mathcal{M}_0$, is a distant third.

\begin{table}
    \caption{Comparison of LOO-CV expected log posterior. The first column contains the model name, the second column is the difference in expected log-predictive density, and the final column lists the difference in standard error. }
    \label{tab:loo_compare}
    \begin{tabularx}{0.95\columnwidth} { 
   >{\raggedright\arraybackslash}X 
   >{\raggedleft\arraybackslash}X 
   >{\raggedleft\arraybackslash}X  }
        \textbf{Model} & \textbf{elpd\_diff} & \textbf{se\_diff} \\
        \hline
        $\mathcal{M}_2$ & $0.0$ & $0.0$ \\
        $\mathcal{M}_1$ & $-205.9$ & $24.7$ \\
        $\mathcal{M}_0$ & $-2217.5$ & $77.0$\\
        \hline
    \end{tabularx}
\end{table}

Based on the DAG and the LOO-CV results, we decided to use model $\mathcal{M}_2$ for future exploration of the posterior predictions.
Worth noting is that $\mathcal{M}_2$ also follows the causal model and, hence, from a causal perspective is the correct model given the assumptions concerning causality in the studied phenomenon.

\subsubsection{Posterior Predictions}

The final step in building a trustworthy model is to visualize posterior predictions and compare them with the collected data points.
The model should fit the trained data, but should also allow some variations in predictions, to avoid overfitting.

\begin{figure}
\centering
\includegraphics[width=0.95\linewidth]{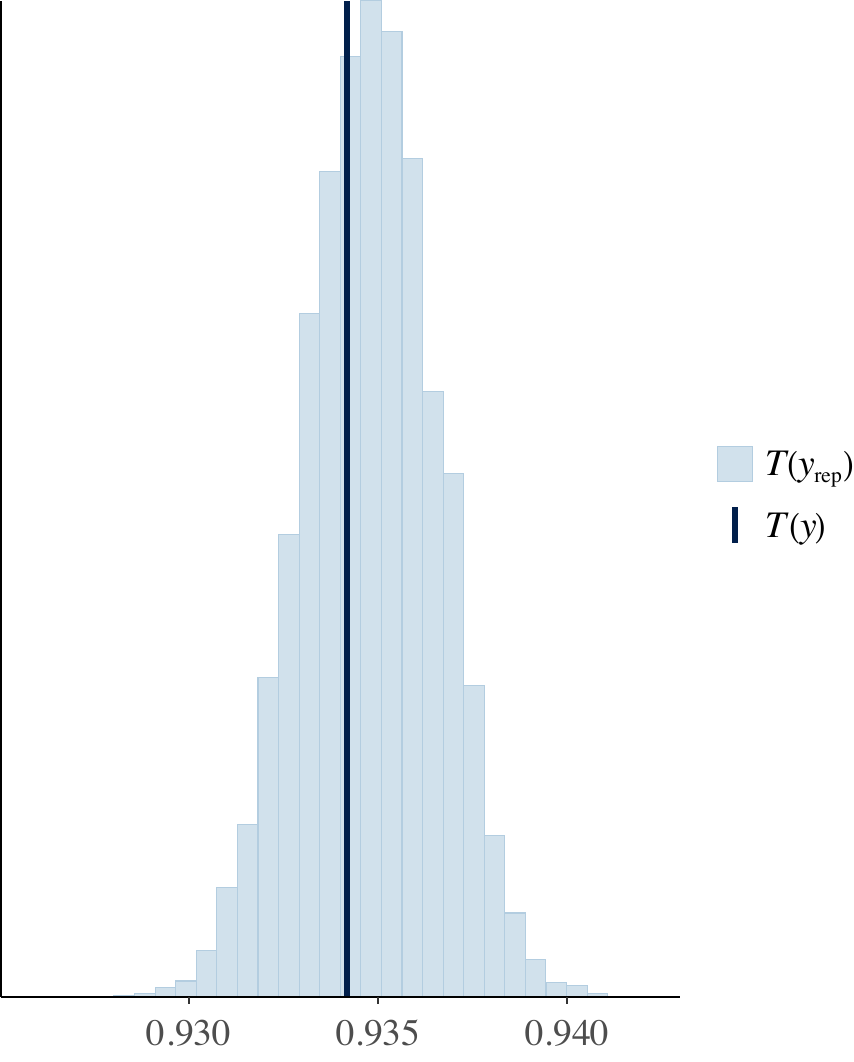}
\caption{$\mathcal{M}_2$ posterior predicted proportion of $0$. $T(y_{rep})$: Model predicted proportions. $T(y)$: Observed proportion of zeros. The observed value lies well within the predicted proportion ($93$\%--$94$\%).
}
\label{fig_post_prop_zeros}
\end{figure}

Figure~\ref{fig_post_prop_zeros} shows the proportion of zeros (no clones added) predicted by model $\mathcal{M}_2$ when given the training data set.
The blue histogram is the predicted proportion of zeros for various samples of the model, and the observed data is the solid dark blue vertical line.
As can be seen in the figure, when seeing the training data set, the model expects between $93$\% and $94$\% zeros.
This fits the observed data (black line $T(y)$) well.
Further posterior predictive checks are available in the replication package.

Once we have a robust model, we can use it to make inferences and visualizations. 
Because the model is multi-dimensional, we typically fixate all but one or two predictors, generate predictions, and then plot the inferences (e.g., expected values and credible intervals) as they vary across the varying predictors.

Further predictions and visualizations are available in Section~\ref{sec:res}.

\subsection{Empirical Validation of $\mathcal{M}_2$}\label{sec:case_study}

Following the methodology from \citet{Runeson2012}, we conducted an embedded case study, where the unit of analysis is a set of components belonging to a particular subsystem developed by a sub-organization in a large software development organization.
We selected components based on availability and organizational structure---the organization was growing during the studied period, and wanted to know how changes to their existing business flow would affect their accumulation of technical debt.

Data collection was initially done via archival analysis (of code repositories and organizational charts), and later augmented by qualitative data from focus group interviews, which was transcribed and subject to open coding.

\anolrevised{We performed four kinds of verification and validation of the model and its results:
\begin{enumerate*}[label=(\roman*)]
    \item we used synthetic data and analyzed whether the different models that we built could recover the used predictors;
    \item our choice of Hamiltonian Monte-Carlo means that the model execution itself will perform fundamental consistency checking (e.g. flagging divergent transitions~\citep{mcelreath2020statistical}), in addition to the standard way to assess model fitting (e.g., rootograms~\citep{kleiber2016visualizing}, for our zero-inflated model);
    \item our choice of using PSIS-LOO (Pareto-Smoothed Importance Sampling,~\citet{vehtari2017practical}) indicates that our model is free from highly influential points that might skew the model results, and finally 
    \item we confronted three development teams, plus line and architect teams, with the results, which allowed us to evaluate whether the model inferences aligned with what they expected.
\end{enumerate*}
}

\subsubsection{Organization and System Characteristics}

The studied system consists of eight components belonging to a subsystem in a large-scale software system.
All components are written in Java and are used in a scalable business processing application that combines event-driven processing with batch-processing tasks.
We studied the components from January 2020 until November 2022, as development ramped up after having been paused for a few years.
During the period under study, the number of contributors grew from 18 in three teams to, at most, 63 in ten teams.
Two teams were disbanded during the study, leaving eight active teams at the end of the study.

The organization initially emphasized ``collective ownership'' of the code repositories, meaning that teams were expected to contribute to all components, and make improvements wherever a developer sees the opportunity.
However, some distribution of repository responsibilities between teams took place during 2021, whereby all development teams were expected to monitor the \textsf{SonarQube}  issues and metrics for one or more allocated repositories.
They were also expected to improve the test coverage in the allotted repositories, unless they were busy working on new features.

As the organization grew, it formed dedicated architect and quality assurance (QA) teams during late 2020.
While the architect team was quite active, both in the production and the integration test code bases, the QA team did not contribute code but instead focused on planning and executing end-to-end tests.

\subsubsection{Data Collection and Exploratory Data Analysis}

The algorithm used for producing the quantitative data is described in Algorithm~\ref{alg:file_change}.
We collected commit data from the \textsf{Main} branch of the studied repositories, as well as organizational data from the team wiki, which contained team composition at different points in time.
In total, ten organization charts were identified for the studied period.
Based on this data, we inferred which team a particular developer belonged to at a particular date. 
We used this affiliation data, together with the author and committer information, to identify the responsible author and the team they belonged at the time of the commit.
Authors lacking known team affiliation (i.e., missing from the organization charts) were modeled as belonging to an ``Unknown'' team---we could have opted to remove or impute these data points, but chose to keep them, as they concerned only 2.0\% of the input data, and had a low impact on the final model.

The \textsf{git} log provided data for each commit, such as date, author, committer, added lines of code ($\mathrm{ADD}$) and removed lines of code ($\mathrm{REM}$).
By merging this with data from the \textsf{SonarQube} analysis, such as McCabe's cyclomatic complexity ($\mathrm{COMPLEX}$), and the number of code clones before ($\mathrm{DUP}$) and after the commit, we constructed a history over which author, and which committer, changed which file, at each point in time.
This was then merged with the organizational data to attribute the change to the team that the author (or committer) belonged to at the time of the authoring (or committing).

In most cases, the author-team and committer-team are the same, and selecting one or the other does not change our conclusions.
But we make the case that, in this organization, having ``free-for-all commit rights,''  it is fairer to attribute the change to the committing team, rather than to the authoring team, as it is the committers team that merges the final change into the master branch.
In organizations with more stringent merge rules or where some automated function merges the change, using the authoring team as the basis of analysis might make more sense.

After completing data collection, we performed exploratory data analysis, to ascertain data quality, and to identify possible patterns.

\subsubsection{Feedback to Organization}

We presented our findings to the studied organization, initially in a meeting with the architect team and the responsible line manager.
In this meeting, we decided to present team-specific findings in meetings with the core teams: Red, Green, and Blue\footnote{At least one developer in each of these three teams were contributing to the product throughout the whole studied period.}.
In the team-specific meetings, we discussed findings particular to the specific team.

We concluded with a summary presentation for the architect team, where we presented the findings together with comments from each core team.
Both the architects and the development teams agreed that the metrics would be useful to gain an insight into team behavior related to clone introductions, but also stated that metrics related to the removal of code clones would be needed to give a more correct picture of the evolution of the code base.

\subsubsection{Modeling Removal of Duplicates}

We concluded the study by making a simple model for the removal of code clones, to see if this model changed some conclusions.

\begin{algorithm}
\DontPrintSemicolon
\SetAlgoLined
\SetKwFunction{AUTHOR}{AuthoredBy}
\SetKwFunction{COMMITTER}{CommittedBy}
\SetKwFunction{ATIME}{AuthoredDate}
\SetKwFunction{CTIME}{CommittedDate}
\SetKwFunction{TEAMATDATE}{FindTeam}
\SetKwFunction{FILES}{Files}
\SetKwFunction{DUPL}{Duplicates}
\SetKwFunction{ABS}{abs}
\SetKwData{filestate}{filestate}
\SetKwData{T}{d}
\SetKwData{AUTH}{author}
\SetKwData{COMM}{committer}
\SetKwData{TEAM}{team}
\SetKwData{DUP}{dup}
\SetKwData{FIXED}{fixed}
\KwData{$C$ commits$: C_i$ precedes $C_{i+1}$}
\KwResult{List of data for changed files}
 \filestate $\gets ()$\;
 \For{$C_i \in C$}{
   $\AUTH \gets $ \AUTHOR{$C_i$}\;
   $\T_{auth} \gets $ \ATIME{$C_i$}\;
   $\COMM \gets $ \COMMITTER{$C_i$}\;
   $\T_{comm} \gets $ \CTIME{$C_i$}\;
   $\TEAM_{a} \gets $ \TEAMATDATE{$\AUTH, \T_{auth}$}\;
   $\TEAM_{c} \gets $ \TEAMATDATE{$\COMM, \T_{comm}$}\;
   \For{$F_{j,i} \in$ \FILES{$C_i$}}{
     $\DUP_{prev} \gets$ \DUPL{$F_{j,i-i}$}\;
     $\DUP_{curr} \gets$ \DUPL{$F_{j,i}$}\;
     $\Delta \gets \DUP_{curr} - \DUP_{prev}$\;
     \lIf{$\Delta < 0$}{$\FIXED \gets$~\ABS{$\Delta$}}
     \lElse{$\FIXED \gets ~0$}
     $\filestate \gets \filestate + (F_{j,i}, \TEAM_a, \TEAM_c, \FIXED)$\;
  }
}
\Return{\filestate}
\caption{Calculating removed clones for changed files in a repository.}
\label{alg:file_change_removals}
\end{algorithm}

We collected data on clone removals using Algorithm~\ref{alg:file_change_removals}, and used an intercept model, structured like model~$\mathcal{M}_0$, with $y$ set to the number of removed clones (\textsf{fixed}), with identical priors as model~$\mathcal{M}_0$, to illustrate the tendency of each team to remove clones.
Results from the model were reported to the architect team in summary format, and are available in the replication package.

Further studies and model designs for clone removals are left as subjects of follow-up studies.

\section{Results}\label{sec:res}

\subsection{Exploratory Data Analysis}

All eight studied components existed at the start of the study and were being developed by three teams (Red, Green, and Blue) at that time.

\begin{figure*}
  \includegraphics[width=\textwidth]{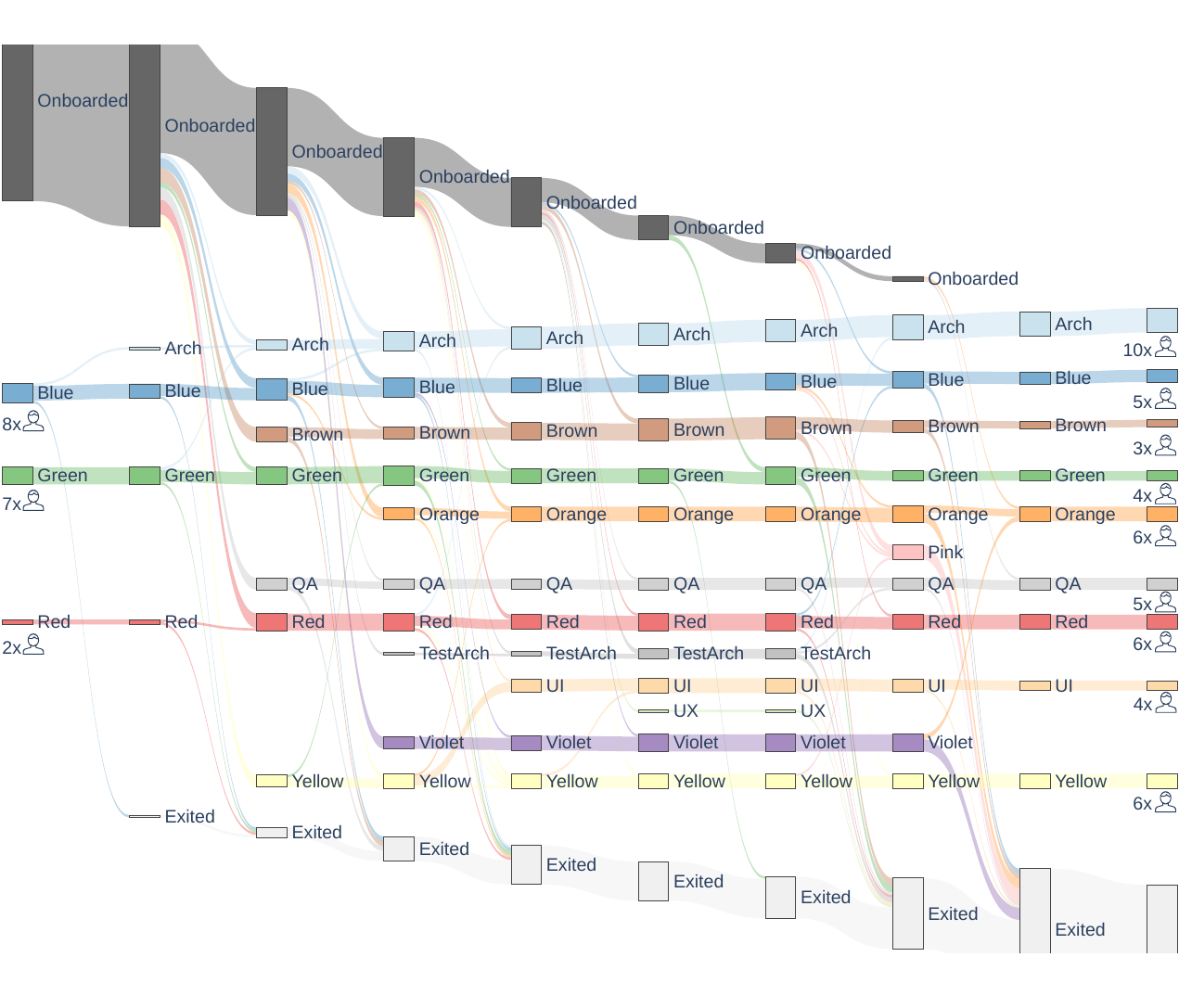}
  \caption{Flow of developers in and out of the organization, and between the different teams.
  Study~start:~Dec.~2019; Study~end:~Nov.~2022
  }
  \label{fig_AffiliationFlow}   
\end{figure*}

At least one developer from each of these teams was part of the studied organization during the whole study period, but all teams changed staffing, to some extent, during the study.
\anolrevised{The number of teams grew both organically (developers switching team affiliation) and via external recruitment.}

Figure~\ref{fig_AffiliationFlow} shows the flow of developers into (\textbf{Onboarded}) and out of (\textbf{Exited}) the studied organization, as well as transfers between teams throughout the studied period.
At the peak, 63 contributors formed eleven teams---including one architect team (Arch), one end-to-end testing team (QA, not contributing to the production-code base), and one UI team (mostly working in non-Java components).
As two of the formed teams were disbanded during the study, at the end of the study eight developer teams, totaling 45 developers, were contributing to the code base, and 5 QA testers were validating the components end-to-end.
In total, 82 developers were onboarded, and 50 left the organization during the study. 

Table~\ref{tab:repo_size} shows the initial ($\textsf{F}_s$) and final ($\textsf{F}_e$) number of Java files, the initial ($\textsf{LOC}_s$) and final ($\textsf{LOC}_e$) lines of code, as well as the number of change events in each repository (\textsf{Chg}).
The components were of different sizes, and changed to varying degrees. In the largest repository, Jupiter, almost nine times more files were changed than in the smallest, Mercury.

\begin{table}
    \caption{Summary code statistics per repository. \\
    $\textsf{F}_s$/$\textsf{LOC}_s$: initial number of Java files/lines of code. \\
    $\textsf{F}_e$/$\textsf{LOC}_e$: closing number of Java files/lines of code. \\
    $\textsf{Chg}$: number of file change events during the study.    
    }
    \label{tab:repo_size}
    \begin{tabular}{l|r|r|r|r|r}
        \textbf{Repository} & $\textsf{F}_{s}$ & $\textsf{LOC}_{s}$ & $\textsf{F}_{e}$ & $\textsf{LOC}_{e}$ & $\textsf{Chg}$ \\
        \hline
        IntTest &  243 &  99310 &  347 & 154637 & 3999 \\
        Jupiter & 1103 & 151628 & 1768 & 219063 & 9413 \\
        Mars    &  413 &  64351 &  729 &  75050 & 2215 \\
        Mercury &  166 &  23437 &  291 &  35597 & 1137 \\
        Neptune &  288 &  38240 &  468 &  66616 & 1801 \\
        Saturn  & 1267 & 149820 & 2157 & 208294 & 6969 \\
        Uranus  &  227 &  26256 &  577 &  60423 & 3083 \\
        Venus   &  198 &  22703 &  482 &  55391 & 2390
    \end{tabular}
\end{table}

\begin{figure}
\centering
\includegraphics[width=\columnwidth]{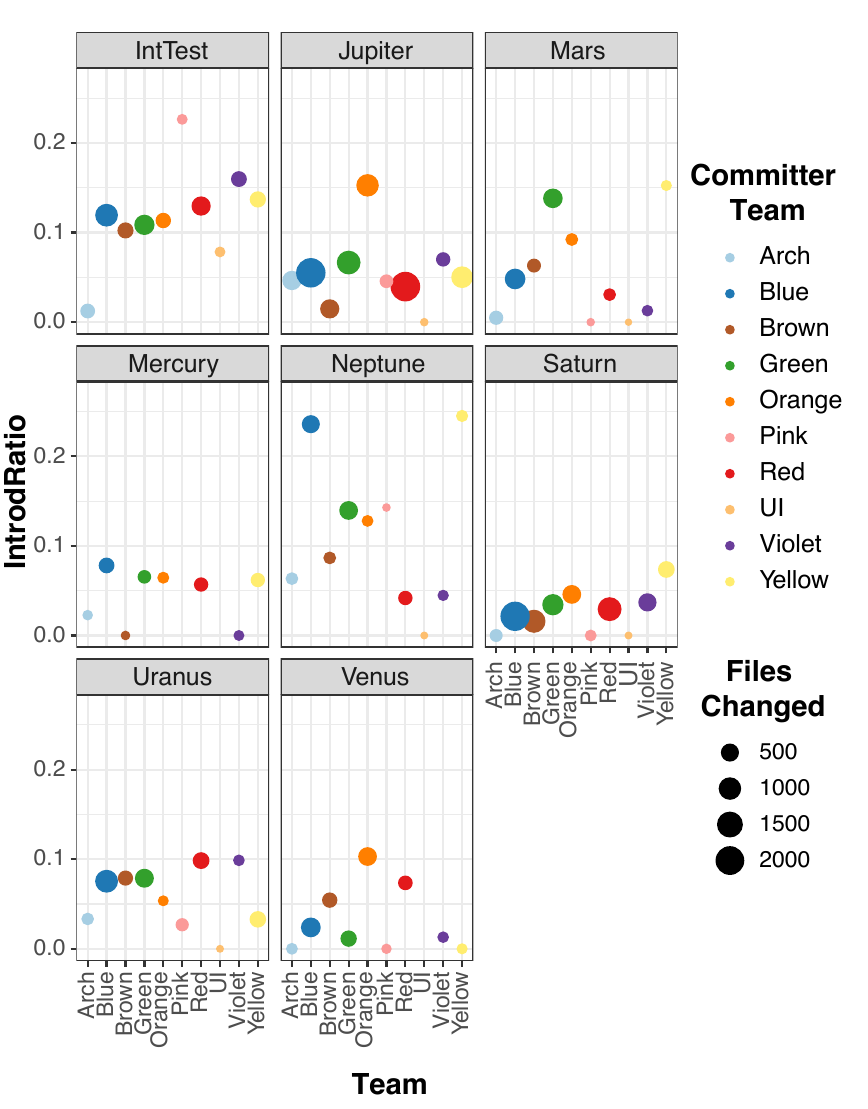}
\caption{Observed proportion of file changes containing one or more new duplicates, per team and repository. The size of the point is proportional to the number of files changed by the team in the repository.}
\label{fig_observed_prop_duplicate}
\end{figure}

Figure~\ref{fig_observed_prop_duplicate} shows the proportion of file changes, per team and repository, that contain at least one new code clone.
The size of the point is proportional to the number of file changes the team has performed in the given repository.
We note that the architect team (``Arch''), though they make fewer changes than the more active teams (e.g., Blue, Green, and Red), are less likely to introduce duplicates.
Furthermore, we note that the repositories have different likelihood of having duplicates introduced.
In the integration test repository, most teams---except Architects---introduce at least one duplicate in every tenth file change, regardless of change size or other predictors.
Some teams make very few changes, in particular Pink, which was active only a few months, and UI, which does not normally work in Java-based repositories.
To avoid selection bias, we chose to keep these data points for model fitting, even though our analysis focused on the more active core teams.

\begin{figure}
\centering
\includegraphics[width=\columnwidth]{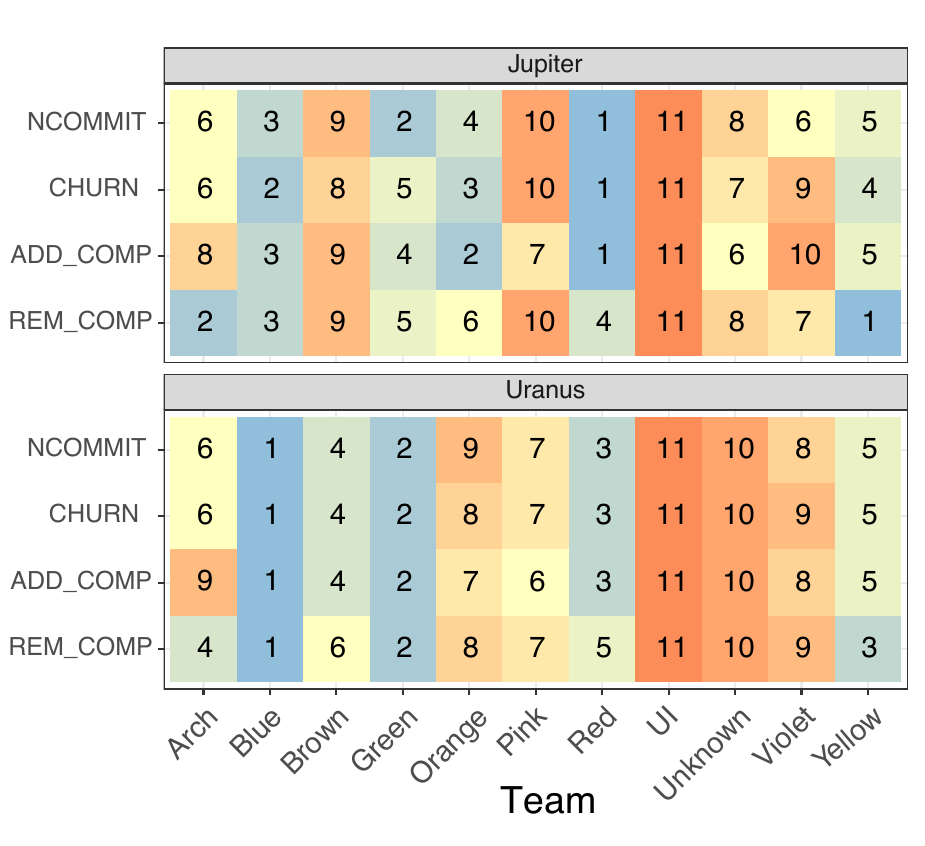}
\caption{OCAM model rank of team contributions to the Jupiter and Uranus repositories.
\textsf{NCOMMIT}:~Number~of~commits. 
\textsf{CHURN}: Sum of \texttt{max(added,~removed)} LOC. 
\textsf{ADD\_COMP}: Sum of added McCabe complexity. 
\textsf{REM\_COMP}: Sum of removed complexity. 
All metrics increase as contributions increase. 
Rank 1 means the highest contribution amongst the studied teams.
}
\label{fig_ocam_jupiter}
\end{figure}

Figure~\ref{fig_ocam_jupiter} shows the OCAM rank for the various teams in the Jupiter and Uranus repositories.
As lower ranks imply higher contributions, we note that in Jupiter, Team Red leads in the number of commits, the total size of changed code (churn), and the added complexity.
Team Blue also ranks high, indicating they have been highly active in this repository.
In Uranus, the roles are practically reversed between Team Red and Blue, and we note that some teams (e.g., Pink, UI, Unknown, Violet) are consistently ranked low.
We used OCAM plots like these to discuss ownership with the teams and contrast these findings with the visualizations from our model.

\subsubsection{Core Teams in the Largest Repository}

\begin{table}
    \caption{Summary statistics for core teams in the Jupiter repository. 
    \textsf{n}: number of changes to files. 
    \textsf{\%dup}: percentage of changes that introduce a clone. 
    $\textsf{C}_{q50}$: median complexity of the changed file. 
    $\textsf{JLINE}_{s}$: initial ratio of authored lines.
    $\textsf{JLINE}_{e}$: closing ratio of authored lines.
    The last row is the total metrics for all teams (including the absolute number of all lines in the Java files).}
    \label{tab:observed_jupiter_changes}
    \begin{tabular}{c | r r r r r}
        \textbf{Team} & \textsf{n} & \textsf{\%dup} & $\textsf{C}_{q50}$ & $\textsf{JLINE}_{s}$ & $\textsf{JLINE}_{e}$ \\
        \hline
        Red           & 2166       & 4.0\%          & 18       & 10.5\% & 17.9\% \\
        Arch          &  623       & 4.6\%          & 16       & ---    & 6.4\% \\
        Green         & 1172       & 6.7\%          & 21       & 12.2\% & 3.8\% \\
        Blue          & 2123       & 5.5\%          & 10       & 15.0\% & 5.0\% \\
        Other         &            &                &          & 62.3\% & 73.4\% \\
        \hline
        Total         & 9413       & 5.9\%          & 13       & 204 kL & 282 kL
    \end{tabular}
\end{table}

Three development teams were part of the organization during the entire studied period, and we also considered the Architect team, which was formed by developers from the Blue and Green teams in early 2020.
Summary statistics for the contributions of the four core teams to the Jupiter repository, the largest in the product, is shown in Table~\ref{tab:observed_jupiter_changes}. Team Red made over $2000$ changes to files in the repository, and Team Green about half as many. 
As part of these changes, the proportion of code authored by Team Red grew from $10.5$\% to almost $18$\%, while the proportion authored by Team Green declined from $12$\% to barely $4$\%.
Although Team Blue made over $2000$ changes, the proportion of code authored by them declined from $15$\% to $5$\%, meaning that they either did significantly smaller changes and/or were consistently changing the same portion of the code in that particular component.
At the end of the study, the four additional development teams had contributed between $0.6$\% to $6.9$\%---summarized in the \textit{Other} row in the table.
About 20\% of the code in the repository had been authored by developers who had left the organization.
About $5.9\%$ of the file changes introduce duplicates in this repository.

\begin{figure}
\centering
\includegraphics[width=0.95 \linewidth]{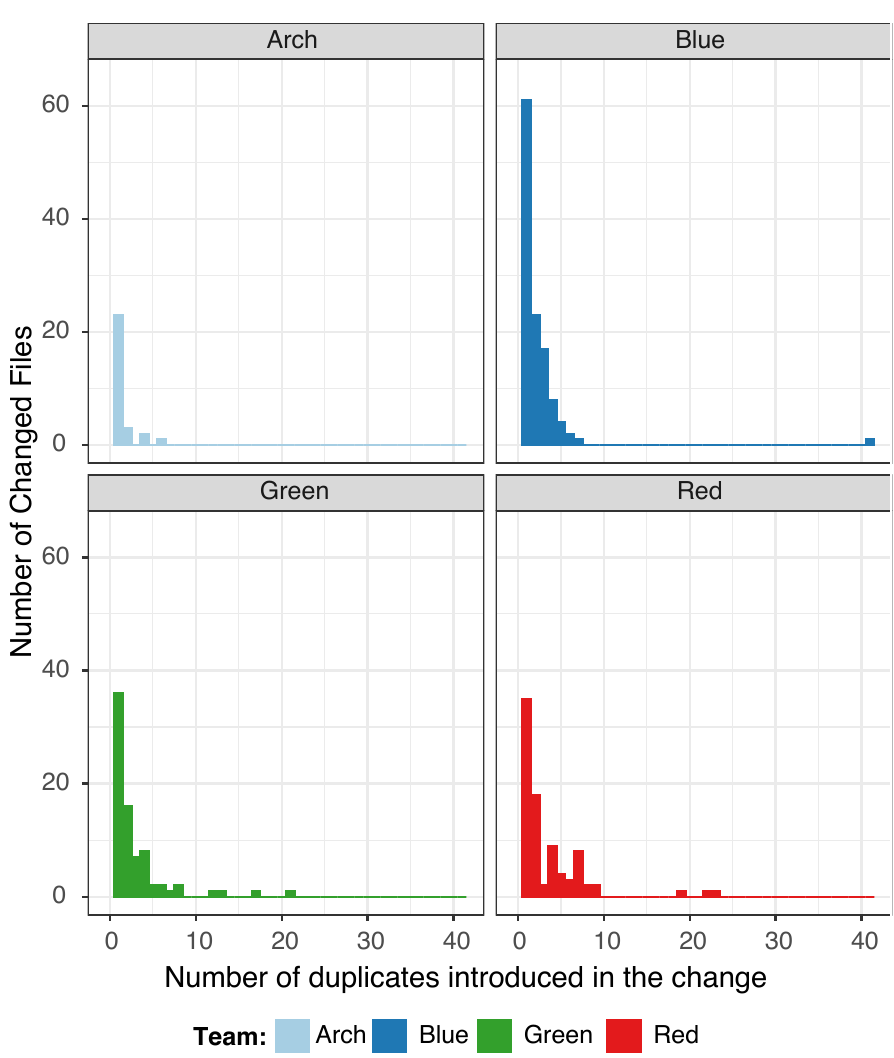}
\caption{Histogram of the observed number of introduced duplicates in the largest repository (Jupiter) by four teams.}
\label{fig_observed_introd_jupiter}
\end{figure}

Figure~\ref{fig_observed_introd_jupiter} show four histograms, one per team, where the $x$-axis represents the number of introduced duplicates, in the Jupiter repository, and the $y$-axis is the number of file changes that add this number of duplicates.
For these teams, eight changes introduce more than ten duplicates (four by Green, three by Red, and one by Blue).
We also note that the architect team introduced fewer duplicates ($23$ single-duplicate-introducing changes, and only a few changes with $2$--$6$ duplicates). 
The Blue team made more than $60$ changes introducing a single duplicate, and they committed one change with more than 40 duplicates, which also was the maximum observed value for this repository across all teams.

The summary data alone shows that teams have different rates of introducing duplicates in the Jupiter repository.
We will explore the differences more using the model in the following section.

\subsection{Fitness of the Model}

\begin{figure}
  \centering
  \includegraphics[width=0.95\linewidth]{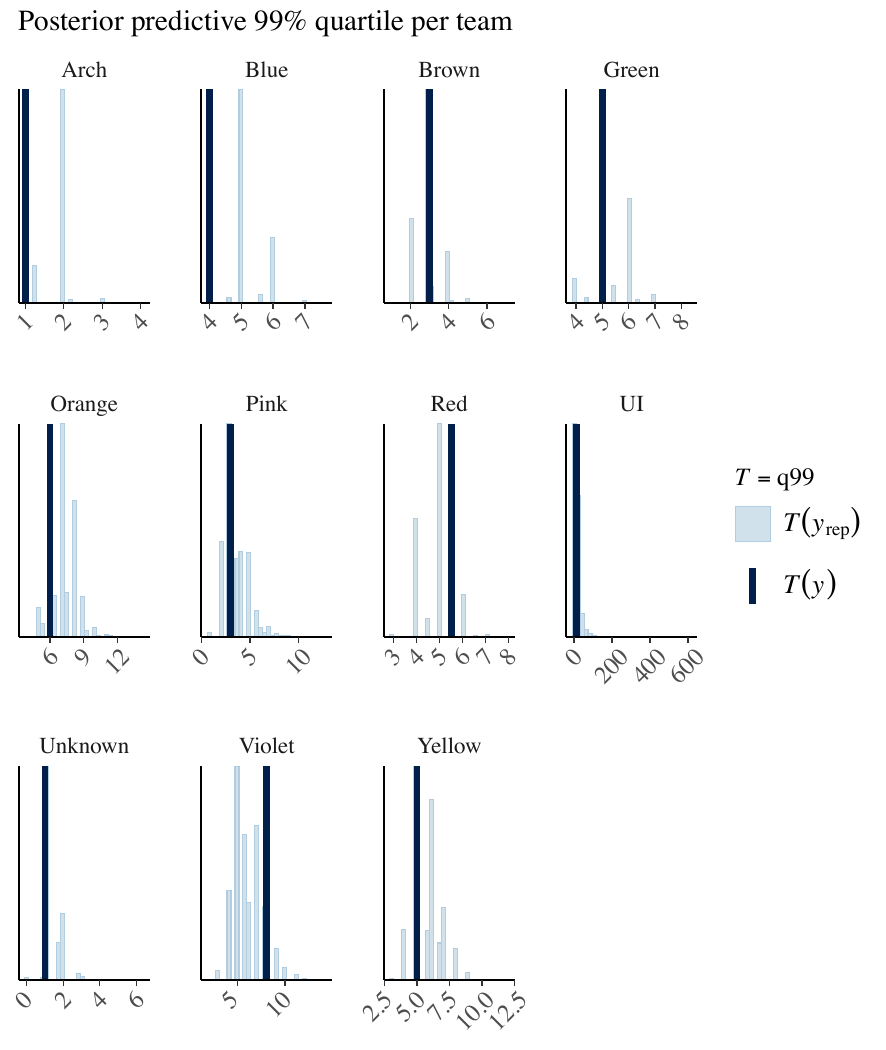}
  \caption{Posterior predicted 99th percentile per team, across all repositories. Note the different $x$-axis scales.}
  \label{fig_99_pct_team}
\end{figure}

Figure~\ref{fig_99_pct_team} shows a posterior predictive visualization displaying the 99th percentile prediction for the various teams across all repositories.
As seen on the respective $x$-axis,\footnote{Note the different scales on the $x$-axes.} most teams are expected to introduce less than 16 duplicates at the 99th percentile.
The exception is the UI team, where some predictions range up into the hundreds.
This is because the UI team only contributed a few changes to a few repositories, causing the model to be more uncertain.
In general, the expected 99th percentile values (light blue histograms) align well with the observed values (solid $T(y)$ line), indicating a good model fit.

More posterior prediction visualizations are available in the replication package.

\subsection{Model predictions}

Given that small changes rarely introduce any duplicates and that all metrics are highly right-skewed, we focus our predictions on choosing either the median (indicated by $Q_{50}$), 95th percentile (indicated by $Q_{95}$), or 99th percentile (indicated by $Q_{99}$) as predictor values.
The observed values for these metrics across all repositories are shown in Table~\ref{tab:observed_parameter_values}.

\begin{table}
  \caption{Summary statistics for file changes in all repositories.}
  \label{tab:observed_parameter_values}
  \begin{tabular}{l | r r r r}
      \textbf{Metric} & $\mathbf{Q_{50}}$ & $\mathbf{Q_{95}}$ & $\mathbf{Q_{99}}$ & \textbf{Max} \\
      \hline
      ADD   &  6 & 143 & 370 & 3772 \\
      REM   &  2 &  92 & 311 & 3413 \\
      COMP  & 16 & 282 & 633 & 1244 \\
      DUP   &  0 &  36 &  99 &  664 \\
  \end{tabular}
\end{table}

\begin{figure}
  \centering
  \includegraphics[width=1.0\linewidth]{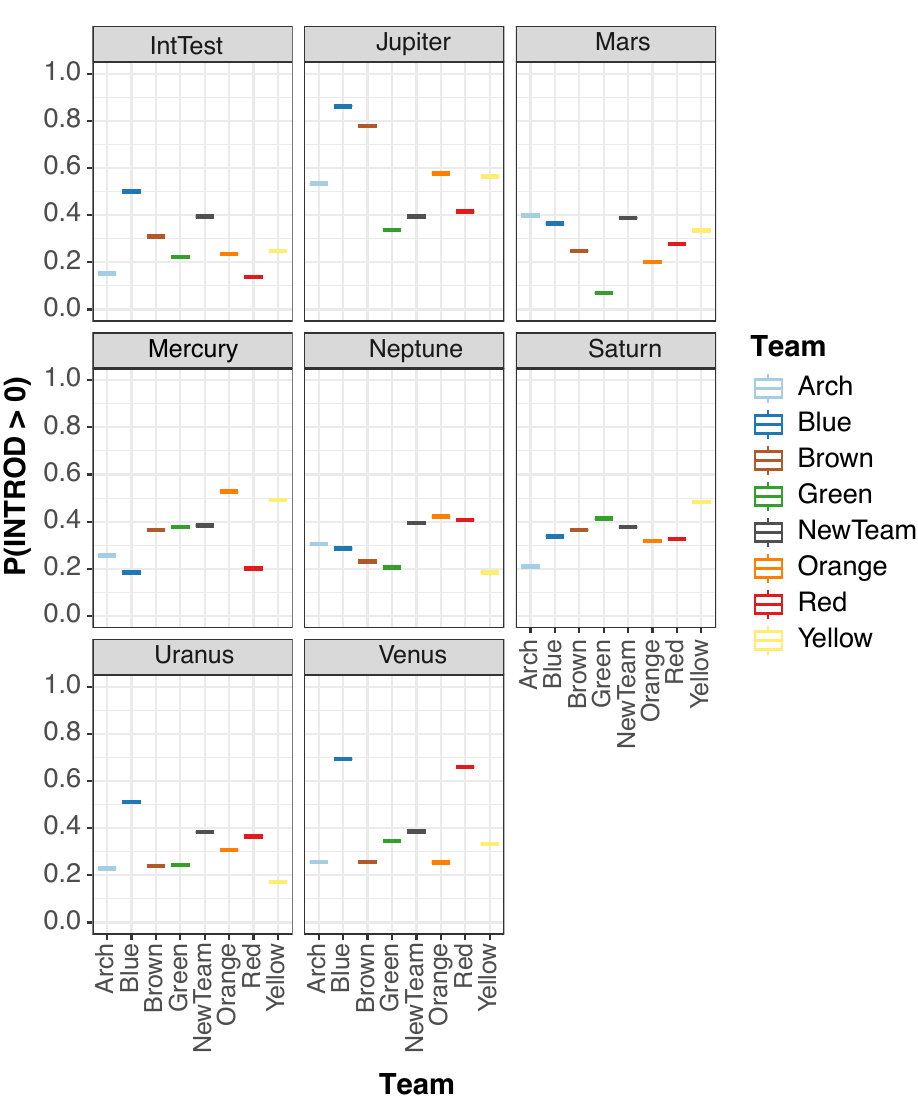}
  \caption{Probability of introducing at least one duplicate for a large change in a complex file.
  ADD~=~370~($Q_{99}$); REM~=~311~($Q_{99}$), 
  COMP~=~282~($Q_{95}$); DUP~=~36~($Q_{95}$)}
  \label{fig_prop_dups}
\end{figure}

Figure~\ref{fig_prop_dups} shows the predicted probability of introducing at least one duplicate for a large change (added and removed lines at their 99th percentile) to a complex file with many duplicates (complexity and existing duplicates at their 95th percentile).
The model predictions are made for the teams active at the end of the studied period, plus a simulated ``Average'' team, which pools information from all teams and repositories.
The different behaviors of the teams are clearly visible in the repositories.
The figure indicates that in many repositories (e.g., IntTest, Saturn, Uranus, Venus), the architect team has a low probability of introducing duplicates---in some cases this behavior is shared with other teams (e.g., Red in IntTest and Green in Uranus).

The largest repository, Jupiter, has the highest probability of teams introducing code clones, in particular by the Blue, Brown, and Orange teams.
For these teams, the model expects more than a 50\% probability that a change with these characteristics will introduce at least one duplicate, whereas the probability for the Green team is around 30\%.
In the second largest repository, Saturn, the only teams that deviate from the norm are the architects, with a \textit{lower-than-average probability} of introducing clones, and Team Yellow, with a slightly \textit{higher-than-average} probability.

\subsection{Model Evaluation---RQ 1, RQ 2}
\anolrevised{Table~\ref{tab:loo_compare} shows that the model that best fits our data, using approximated leave-one-out cross-validation (LOO-CV), is~$\mathcal{M}_2$, described in Eqs.~\ref{eqn_zinb_m2}--\ref{eqn_zinb_m2_priors}.}
This is a Bayesian MLGLM, with four numerical and two categorical predictors, with varying slopes and intercepts.
To assess model convergence, we used standard diagnostic checks~\citep{mcelreath2020statistical,vehtari2017practical,gelman2020bayesian}, such as $\widehat{R}$, $n_{eff}$-ratio and Pareto-$k$ value approximation, and to assess model fit, we used the rootogram method recommended by~\citet{kleiber2016visualizing}.
Detailed diagnostics are available in the replication package~\citep{epkanol_2024_11357296}.

\subsection{Predictions of the Estimated Number of Introduced Duplicates }

Using $\mathcal{M}_2$, we can make predictions using the posterior distribution, to visualize and summarize results.

\begin{figure}
  \centering
  \includegraphics[width=\columnwidth]{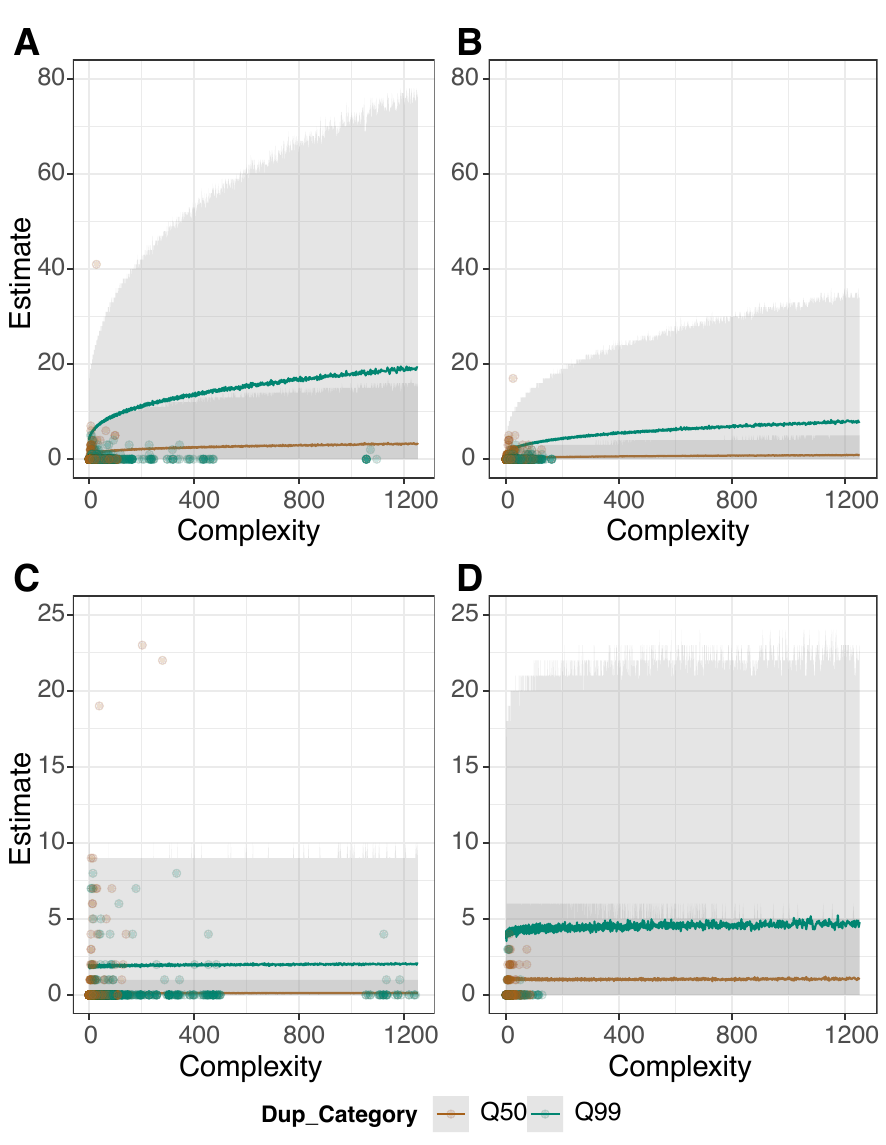}
  \caption{Estimated number of introduced clones for a large change, by complexity and existing duplicates.
  $Q_{50}$:~DUP~=~0; $Q_{99}$:~DUP~=~99
  A)~Team~Blue~in~Jupiter; B)~Team~Blue~in~Uranus;
  C)~Team~Red~in~Jupiter; D)~Team~Red~in~Uranus;
  ADD~=~143~($Q_{95}$); REM~=~92~($Q_{95}$);
  observations as points, regardless of change size. 
  The shaded area represents the 89\% credible interval. 
  Team Red is largely indifferent to existing complexity. 
  Both teams are sensitive to existing duplicates, and are expected to introduce fewer clones in repositories that they have chosen to be responsible for.}
  \label{fig_condeffect_by_complexity}
\end{figure}
Figure~\ref{fig_condeffect_by_complexity} shows how the estimated number of introduced duplicates varies for a large change, made by two teams (Blue and Red) in two repositories (Jupiter and Uranus).
The $x$-axis represents the complexity of the changed file, and the different colored lines represent the expected number of introduced clones, depending on whether the file had any existing duplicates or not.
Both added and removed lines are set to their respective 95th-percentile value.
The shaded areas represent the 89\% credible interval for predictions---all areas start at 0, and the larger the number of existing duplicates, the larger the credible interval.
The figure also contains observed data points for the respective teams, regardless of the size of the change, but colored by whether the file had duplicates or not.
\begin{tcolorbox}[colback=gray!5!white,colframe=gray!75!black]
  \textbf{\textit{Complexity matters---but not for all}}\\
  As is shown in Fig.~\ref{fig_condeffect_by_complexity}, the expected number of duplicates for Team Red (bottom row) does not depend on the existing complexity of the changed file.
  The other teams show varying strengths in the association between existing complexity in the file and the expected number of introduced clones. 
  The preexisting number of clones plays a role in all teams---files without any existing clones are unlikely to see a large number of introduced clones, and the larger the number of existing clones, the larger the expected number of newly introduced clones.
\end{tcolorbox}

During 2021, Team Blue (top row) chose to be responsible for Uranus (B), while Team Red (bottom row) chose Jupiter (C).
As the figure shows, both teams are expected to add around half as many duplicates in the repository they assumed ownership of, relative to the other.

\begin{tcolorbox}[colback=gray!5!white,colframe=gray!75!black]
  \textbf{\textit{(Some) owners are different}}\\

  As shown in Fig.~\ref{fig_condeffect_by_complexity}, assumed ownership of a component does seem to impact the rate of clone introductions.
  In this study, teams were allowed to self-select which components to be responsible for, from the perspective of reducing \textsf{SonarQube} violations and increase test coverage.
  There was no formal team ownership (such as requiring sign-off on commits or patch sets) in place.
  As illustrated in the figure, both Team Blue and Team Red appear to introduce about half as many duplicates in the components they chose to be responsible for, relative to the other.

\end{tcolorbox}

Given what~\citet{dietz2003struggle} say about the importance of \textit{monitoring of the common resource} to efficiently govern its usage, we postulate that our model, and in particular the posterior predictions and visualizations would be a useful tool for architects and team leaders to judge the adherence of individual teams to the agreed code duplication policies.
Unlike standard statistical summaries (such as Fig.~\ref{fig_observed_prop_duplicate} and Fig.~\ref{fig_observed_introd_jupiter}), which does not take complexity and existing number of clones into account, our model adjusts its predictions to the circumstances of each file change.
Thus, it has the potential to be more precise, and therefore, more trustworthy, for developers and architects alike.

\subsection{Comparing with the Average}

\begin{figure}
  \centering
  \includegraphics[width=\linewidth]{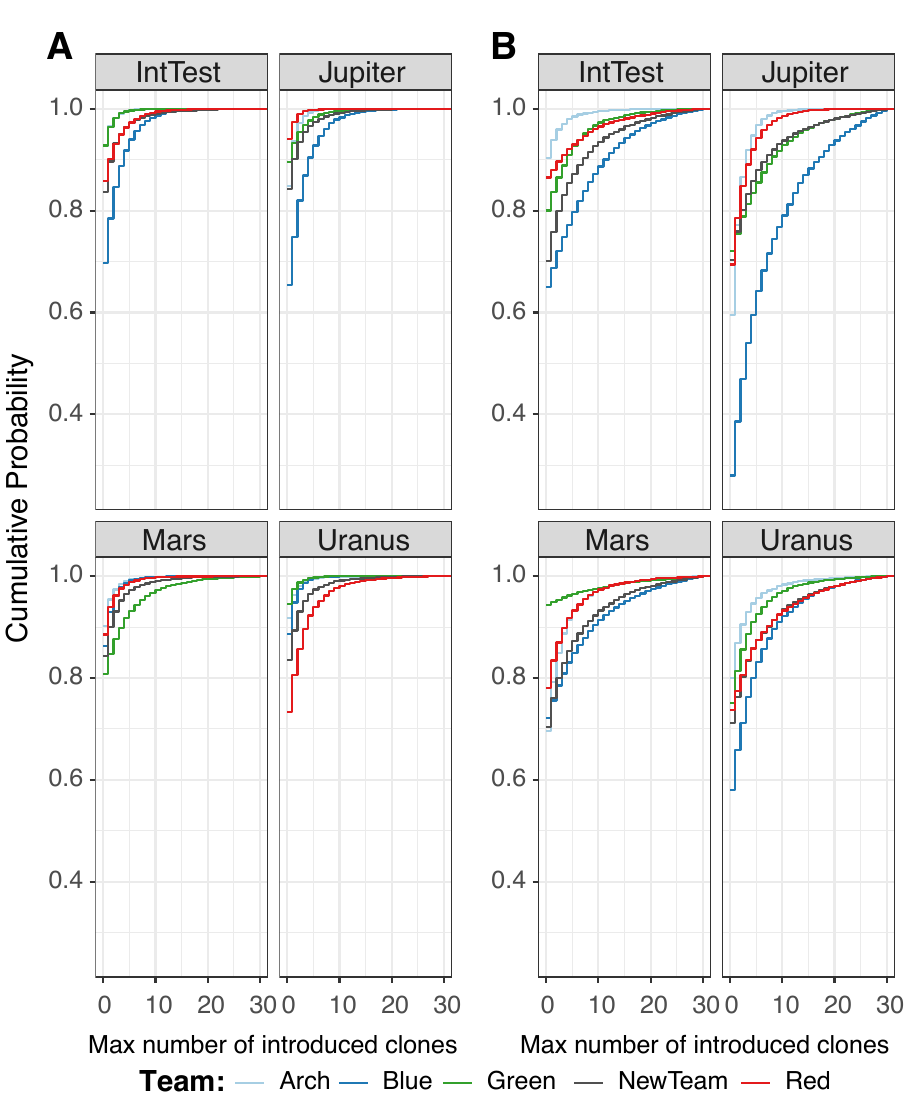}
  \caption{Cumulative probability of introduced code clones for four existing and a simulated new team, making a large change; ADD~=~143~($Q_{95}$); REM~=~92~($Q_{95}$); in four repositories; \textbf{a)} in a file of median complexity lacking existing duplicates;~COMP~=~16~($Q_{50}$); DUP~=~0~($Q_{50}$), and \textbf{b)}~in a complex file with many existing duplicates; COMP~=~282~($Q_{95}$); DUP~=~36~($Q_{95}$).}
  \label{fig_predicted_q95}
\end{figure}

Using the model, we can simulate how a new team (i.e., a team that has not yet produced any data points) is likely to behave, based on the behavior of existing teams, using a concept called ``partial pooling'' in Bayesian contexts. 
This can give interesting comparisons. 
Figure~\ref{fig_predicted_q95} shows the expected cumulative probability of introducing duplicates for four existing teams, as well as the population average (``NewTeam'') in four repositories, for a fairly large change (added and removed lines to their 95th percentile). 
The $x$-axis shows the maximum number of introduced duplicates and the $y$-axis shows the probability of seeing the corresponding $x$ number of introduced clones, or less.
In Figure~\ref{fig_predicted_q95}, part $A$, the complexity and existing number of duplicates are set to their median values (16 and 0, respectively), indicating a large change in a median complex file. In Figure~\ref{fig_predicted_q95}, part $B$, all values are instead set to their 95th percentile values (complexity 282 and 36 existing duplicates), indicating an equally sized change, in a more complex file.
Based on the figure, we can draw several conclusions:

\begin{itemize}
    \item ``NewTeam'', the population average---being the average---can be used to indicate whether a team is more or less likely to introduce duplicates than the average team. 
    We see that the placement of teams, relative to the population average, varies across repositories, and depending on complexity.
    For a change in a median file in Uranus, Team Red is below average, whereas they are close to the average for a complex change.
    The opposite is true for Team Blue, who are more affected by complexity.
    \item Team Blue, as stated, behaves below average for both integration tests and Jupiter, the largest repository, where they have slightly less than 30\% probability of avoiding introducing duplicates in a complex file.
    In Mars and Uranus, changing median complex files lacking duplicates, they behave better than the average team.
    \item Team Red, in general, behaves like the average team, or better, with the exception of changing median complex files in Uranus.
    The model expects the probability of introducing a duplicate in Uranus to be similar regardless of complexity and whether there are duplicates or not (though the slope is different, so the number of duplicates introduced would be expected to be higher in a more complex file)
    \item Team Green in the Mars repository, is very unlikely to introduce clones in the complex file, but in the median file, they behave slightly worse than average (though the probability of avoiding clones is still around 80\%). 
    The reason for this is to be found in the data---the majority of cases where Team Green has introduced clones in Mars have been in files lacking any existing duplicates, and they have never introduced a clone in a file with more than two existing duplicates.
    Thus, the model infers that when Team Green makes a change in a complex file with existing duplicates, they are unlikely to introduce any clone.
    \item Architects are, in general, less likely to introduce clones, particularly in the integration test repository, where the model infers around a 90\% chance of avoiding clones in a complex file.
    One exception is the probability (around 60\%) of keeping zero clones for a change in a complex Jupiter file---but the probability of more than one clone drops off more quickly than the average team.
\end{itemize}

\begin{tcolorbox}[colback=gray!5!white,colframe=gray!75!black]
  \textbf{\textit{Pathbreakers vs.\ caretakers}}\\
  When confronted with the findings from the collected data, both the Architect team and Team Blue stated that the reason for the aberrant behavior of Team Blue in the Jupiter repository was that they handled ``the most difficult tasks'', and during the studied period they were actively working on rearchitecting the main business flow, largely encapsulated inside the Jupiter repository.
  Thus, they could be seen as ``pathbreakers'', versus the more careful and deliberate ``caretakers'', i.e., the Red and Green teams.
\end{tcolorbox}

\subsection{Teams' Feedback---RQ~3}

While the organization lacked any formal ``gatekeeper'' form of ownership of the repositories, the teams did state that they, during 2021, had distributed the repositories between themselves to manage \textsf{SonarQube} violations and improve test coverage.
Teams were allowed to self-select repositories to care for, and based on the (LOC) size of each repository, one or more repositories were chosen.
While Team Red had responsibility for one repository---Jupiter, the largest, Team Green had seven, among them Mars, and Team Blue had two, one of which was Uranus.
The Architect team was not allocated any repository, as they had the overall product structure responsibility.

Figure~\ref{fig_predicted_q95} shows that the Architect team in the integration test repository has a very low probability of introducing clones, relative to the normal development teams.
They attributed this to the fact that ``we make changes to this repository only when we change various interfaces of the product''.
Looking at the data, this statement is partially correct---the Architect team does introduce fewer new files (14 of 241, $\approx5.8\%$) than the Blue (93\slash 1030, $\approx9.0\%$) and Brown (24\slash 332, $\approx7.2\%$) teams.
However, many teams are less active, and Team Red (another core team) introduced 30 new files out of 617 changes ($\approx4.9\%$).

The Blue team in the Jupiter repository shows a stark difference from the other teams.
This repository is the largest in the product, where most of the business logic is contained.
When we discussed this with the team, we got a partial explanation:
``We were in Proof-of-Concept mode, no process was followed. 
Lots of code was copied across repositories$\dots$ 
And after that, there was a reset, where 1.5 years of work was almost null and void$\dots$''
The Architect team backs this up:
``It was a huge feature, a twelve--fourteen-kind-of-sprints feature, changing almost every aspect of the two main business flows. 
Because of this, they might have been replicating parts of it, causing duplicates''.

As depicted in Fig.~\ref{fig_predicted_q95}, Team Green behaves close to the population average in most repositories.
However, in the Mars repository, they behave significantly differently, having a $90$\% chance of avoiding introducing a duplicate when changing a complex file, compared to the population average of around $75$\% (a figure that is also close to the Red and Blue team behavior).
Team Green responds to this fact with:
``In Mars, we have achieved kind of a standard of code, whereas in Jupiter we are introducing diverse code, which results in bad coding practices sometimes''.
This indicates that this team is familiar with Mars, while being more foreign to Jupiter.
The other teams do not appear to be as familiar with Mars as Team Green.

All teams state that they think the statistics and visualizations give important insights into team behavior, but also state that a fuller picture would be gained by also modeling how teams improve code by removing duplicates.

\subsection{Model for Removal of Code Clones}

As the teams indicated that the removal of clones by teams in different components should also be considered, we tried to model clone removals via the same method as the introductions.
We collected data on clone removals according to Algorithm~\ref{alg:file_change_removals}, and then fit the simplest, intercept-based model onto the collected data to get the average behavior of the team in each repository without considering any predictors.

To construct model $\mathcal{M}_3$, we used Equation~\ref{eqn_zinb}, with $y$ representing the number of (zero or more) removed code clones.
The parameters of the zero-inflated negative Binomial likelihood are defined according to Eqs.~\ref{eqn_zinb_m0_mu_xi}--\ref{eqn_zinb_m0_team}.
As the prior predictive checks showed reasonable values, we used the same priors (defined by Eq.~\ref{eqn_zinb_m0_priors}) for model~$\mathcal{M}_3$ as for model~$\mathcal{M}_0$.

\begin{figure}
  \centering
  \includegraphics[width=\linewidth]{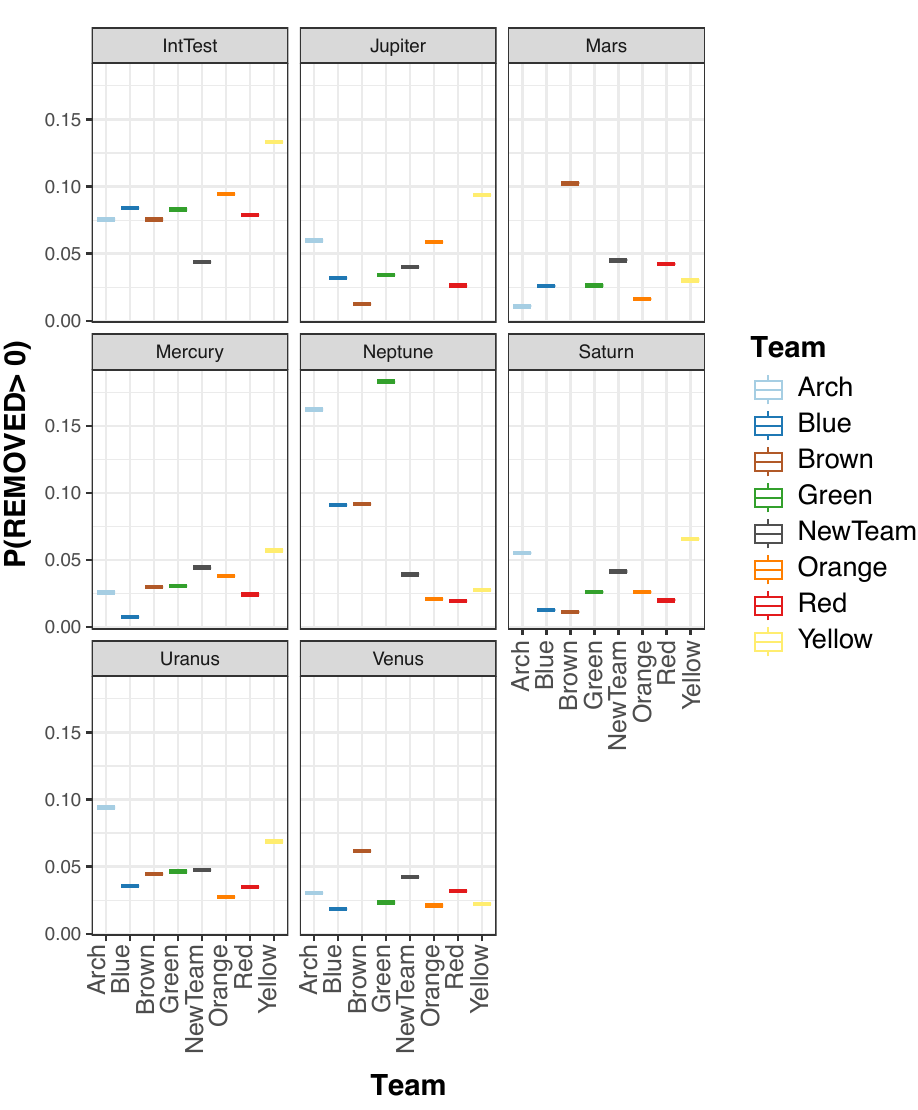}
  \caption{Probability of removed clones, based on an intercept-only model.}
  \label{fig_predicted_removals}
\end{figure}

Figure~\ref{fig_predicted_removals} shows the probability of removing one or more duplicates for the eight currently active teams in the various repositories.
A few teams and repositories stand out, such as the Architects and Team Green in Neptune, Team Yellow in integration tests, and Team Brown in Mars.
But the general pattern is that the teams show quite similar behavior, and the simple intercept model (not considering any predictors) expects, at most, a $5$--$10$\% chance of a team removing a clone, regardless of which repository.

\section{Discussion}
\anolrevised{
In this section, we separate the discussion of the general findings from the case-specific ones.
}

\subsection{Monitoring Code Clone Introduction Trends}
\label{sec:monitoring_trends}

\anolrevised{
Relative to summary statistics, such as the proportional data in Fig.~\ref{fig_observed_prop_duplicate} and Table~\ref{tab:observed_jupiter_changes}, our Bayesian model allows instant ``grading'' of a contribution.
Using the tools that generated Fig.~\ref{fig_predicted_q95}, each file change in a merge request can be automatically rated, both in comparison to the average team and relative to the typical behavior of the developing team.
Given input data (i.e., the number of added and removed lines, the existing complexity and number of duplicates in the file, and the author team), the model would output the probability of seeing \textit{y} new clones.
This could be integrated into automated review tools, such as \textsf{SonarQube} or \textsf{Gitlab} pipelines.
Providing this sort of feedback \textit{before} merging to the main branch would allow the team to proactively affect their ``clone introduction probability,'' before the next iteration of the model is built.
}

\anolrevised{
Statistical summaries such as the ones in Fig.~\ref{fig_condeffect_by_complexity} and Fig.~\ref{fig_predicted_q95} could be used by architects or experienced developers to ascertain how newly formed or onboarded teams fare, relative to ``the average team'' and detect when they might have a different behavior than what could be expected.
This should allow an objective assessment of how teams behave from a clone-introducing perspective, relative to the average team, in different components.
It is important to note that, as the averaging occurs across the entire population of developer teams, there will always be \textit{some} teams that are better, and \textit{some} that are worse than the average.
But the model also allows an objective assessment (with uncertainty estimates) of \textit{how much} better or worse, each team is, relative to the average team.
This should allow the architects to gain insights that can be used to initiate discussions about how each team are performing in the different components of a subsystem, which in turn could be used to identify components where more architectural support is warranted.
}

\paragraph{Choice of Grouping Level for Code Clone Introduction Modeling}

Relative to earlier software engineering studies, which usually mapped out ownership based on the number of changes individuals make to distinct files~\cite[e.g.][]{bird2011don,AVELINO201914,foucault2015usefulness}, our study aligns more with the OCAM~\citep{zabardast2022ownership} way of inferring teams based on the author or committer of a particular file change.
In the studied organization, teams were responsible for what they committed, including reviewing and testing the change, and we validated this inference with the studied organization.
We attribute the team affiliation based on where the committer was located at the time of the change.
This means that the ownership of components will fluctuate over time, as people change teams~(as seen in Fig.~\ref{fig_AffiliationFlow}).

\anolrevised{
  In other organizations, like Open-Source projects with loosely affiliated teams and occasional corporate contributors, different methods might be applied to assign team labels to file changes.
  In a typical successful Open-Source system\footnote{\url{https://docs.linuxfoundation.org/lfx/project-control-center/v2-latest-version/reports/health-metrics/code-contributions}}, committing to active branches is often reserved to a select set of \textit{Committers}, while \textit{Reviewers} are responsible for reviewing incoming code changes, and \textit{Maintainers} are responsible for the coordination of contributions, releases, and the overall project direction.
  The Linux kernel project\footnote{\url{https://docs.kernel.org/process/submitting-patches.html}} allows different headers in its patches (e.g., \texttt{Co-developed-by:}, \texttt{Acked-by:}, \texttt{Signed-off-by:}, \texttt{Reviewed-by:}), which should be used to identify contributing individuals.
  Depending on the question you are trying to answer (e.g., ``In general, how do component~\textit{C} patches, originating from Organization~\textit{X}, compare from a clone perspective to those written by~\textit{Y}''), one or more of these attributes could be used to determine the contributing team, in addition to the \textsf{git} standard \texttt{Author:} or \texttt{Committer:} attributes.
  This might also require updates to the causal model underlying the Bayesian model.
}

\anolrevised{
The Bayesian model is agnostic to the choice of grouping level; it will use whatever groups (labels) the data contains.
However, too wide groups (e.g., sub-organizations, geographical sites, or countries) might produce more diffuse, less-actionable results, since the findings might be harder to connect to team practices and team behavior.
Conversely, groupings that are too fine-grained (e.g., on individual author level) might produce groups with too few data points to make reliable predictions.
Additionally, attributing a change to a single author would introduce bias by disregarding contributions made by others, for instance, due to pair programming, reviewing, or peer testing.
}

\paragraph{Capturing the Diversity of Team Behaviors}

\anolrevised{
  The visualizations of different teams' probability of introducing one or more duplicates (e.g.~Fig.~\ref{fig_predicted_q95}) can be used to discuss the causes of clone introductions with the team.
  Team Blue was, by architects and themselves, seen as ``pathbreakers,'' who were often tasked to perform the more complex tasks (such as architectural changes), relative to the more ``caretaking'' Team Red (who, in turn, were standing out in that they were, in most components, unaffected by existing complexity when introducing clones).
  Thus, the visualizations can guide architects to understand, and initiate discussions with, different teams, in a language that they understand.
  Both Team Blue and the Architect team agreed that neither a formal process nor \textsf{SonarQube} rules were followed while Team Blue was changing the architecture as a ``Proof of Concept.''
  This is the likely reason for them being much more likely to introduce clones in the largest repository in the product (and also, to a lesser extent, in the other repositories).
}

\paragraph{Comparing With the Average, Rather Than Between Teams}
Figure~\ref{fig_predicted_q95} illustrates how a Bayesian model can simulate a new team, based on the pooled population average.
The behavior of the ``generic new team'' is based on the partial pooling of existing teams, with the influence of individual teams proportional to how much they have contributed to the various repositories.
These predictions can be used to monitor the progress of both new and existing teams and detect behaviors that could be improved, or competencies that should be widened.

\anolrevised{
Relating to the findings made by~\citet{dietz2003struggle} to avoid the ``Tragedy of the Commons'', we postulate that our model, and its visualizations can serve as a useful tool to \textit{monitor} (or act as a \textit{quality gateway} for) both newly onboarded, as well as more experienced, teams making contributions to existing components.}
\anolrevised{The architect team also indicated that the visualizations could improve \textit{communication} about how and why teams introduced code clones in different components.
Comparing with the average (rather than pitting one team against another) might also improve the \textit{social networks} between architects and development teams.
However, seeing any long-term effects requires more longitudinal observations, which we defer to future studies on this subject.
}

\subsection{Case-specific findings}

\anolrevised{
In the following sections, we discuss the major findings highlighted by the model, as well as how the teams reacted to these findings.
}

\paragraph{Architects are Active and Also Behave Differently}

In this organization, the architect team is active in all repositories, and employs a different behavior than the regular development teams.
They have the highest median number of removed lines (7), which is higher than their median number of added lines (3).
Most other teams have higher median added lines (ranging from 3 to 20) than median removed lines (ranging from 0 to 4).
Figure~\ref{fig_predicted_q95} shows that the behavior is especially pronounced in the integration test repository, where the architect team is between $5$--$25$\% less likely to introduce duplicates in a complex file than the other teams.
The architect team claims that this is because they are mainly making changes due to added or changed external interfaces.

\paragraph{Teams React Differently to Existing Complexity}

As visualized in Fig.~\ref{fig_condeffect_by_complexity}, Team Red is largely indifferent to existing complexity with regard to the rate of clone introduction.
For all repositories but one (Venus), this team seems to be unaffected by complexity, although the rate of clone introduction will vary across repositories (as shown in Fig.~\ref{fig_condeffect_by_complexity}, parts C and D).
Team Blue, on the other hand employs the more common behavior, where the likelihood of introducing code clones increases, as the complexity of the file grows.
\anolrevised{
Team Red stated that they are careful to avoid introducing new clones, and that they have started fixing old \textsf{SonarQube} violations as they go along with their contributions.
They have not contributed much code to the Venus repository, which could be one reason for why the model assumes they behave like an average team there.
}

\paragraph{Ownership Matters}

In Fig.~\ref{fig_condeffect_by_complexity}, Team Blue is responsible for Uranus (B) and is less likely to introduce clones here than in Jupiter (A).
The reverse is true for Team Red, who is responsible for Jupiter (C), but not for Uranus (D).
In Fig.~\ref{fig_predicted_q95}, Team Green is responsible for Mars, and has the lowest likelihood (about 10\%) of introducing clones of all teams, when changing complex files.
Initially, the studied organization employed total collective ownership.
However, this changed in 2021, when the teams were allowed to \textit{self-select} which components they should care for by fixing \textsf{SonarQube} issues and improving test coverage.
To some extent, this can explain the behavior---teams would likely select components where they ``felt at home'' when they allocated team ownership.
Because of this, we cannot claim causality, but we can state that for these three teams, ownership allocation was reflected in how they behaved in the respective components.

\anolrevised{Given that most Open-Source projects also employ self-selection, it would be interesting to evaluate whether this finding applies also to such organizations, and how it varies for corporate contributors.
To make such an assessment, in addition to the Bayesian model and a tool such as SonarQube, the following would be needed:
\begin{enumerate*}[label=(\roman*)]
  \item the full source code, including commit history (to build the historical data needed for predictions) of the project you want to analyze;
  \item how to map contributions using labels such as \texttt{Author:}, \texttt{Committer:} or any of the ones mentioned in Chapter~\ref{sec:monitoring_trends}, into the team (or other grouping level of interest) responsible for the contribution;
  \item how the responsibilities of these groups are mapped to the \textit{Contributor}, \textit{Committer}, \textit{Reviewer}, and \textit{Maintainer} roles for the various components of the system under study. We would assume that committers (and possibly maintainers) would show the strongest ownership behaviors, followed by reviewers, and last contributors. If our assumption holds, for a given component, we would see groups with the committer or maintainer role behave similarly to Team Green in Mars or Team Red in Jupiter, relative to how they behave in other components, where they are merely contributors.
\end{enumerate*}
Unfortunately, we have to defer the studies of such Open-Source systems to a follow-up study.
}

\paragraph{Do Not Forget Removals}

When discussing our findings with the teams, they often asked about clone removals.
To illustrate this, we also developed a simple (intercept-based) model for how teams remove clones in the various repositories.
This is illustrated in Fig.~\ref{fig_predicted_removals}, where we see the overall clone removal probability for various teams, in the various repositories.
There are some teams that ``stand out'', such as Team Yellow in integration tests, and the Architects and Team Green (and to a lesser extent Team Blue and Team Brown) in Neptune.
Overall, this model illustrates that most teams behave quite similarly regarding clone removals.
Further research in this area is deferred to a follow-up study to explicitly model removals with a different DAG and possibly different predictors. 

\section{Threats to validity}

We structure our threats to validity according to the four different angles recommended by~\citet{Wohlin2012experimentation}.

\textbf{Construct validity} concerns whether the studied measures reflect what the researcher had in mind, and what is stated in the research questions.
We base our study on a causal DAG and attribute latent constructs (such as team cleanliness and knowledge of components) to quantitative data (such as introductions of code clones).
To aid interpretation of the results, and improve construct validity, we presented the findings to four teams, and included their feedback in this paper.

Changes to source code are made by individuals, based on more or less feedback from other individuals.
In this study, we attribute the \textit{committing} individual, at the time of the commit, to the associated team according to the organization chart, and use this information to make inferences about general team behavior.
Several possible threats arise from this attribution:
\begin{enumerate*}[label=(\roman*)]
    \item modern software development methods (e.g, pair or mob programming), and software version control systems (e.g., Gerrit\footnote{\url{https://www.gerritcodereview.com/}} and GitLab\footnote{\url{https://about.gitlab.com/}}) allow multiple developers to collaborate on a \textit{patch set}, before one person finally merges it, as one or more commits, into the master branch; 
    \item individuals do not necessarily represent the team they are working in---misrepresentations are possible, for instance by ``lone wolf development,'' or by individuals working closer to other teams than their officially assigned;
    \item the committing individual might have no relation at all to the actual changes---in some organizations, the commit function is outsourced to an automated function, that merges the code once all required tests pass.
\end{enumerate*}
Regarding threat (i), we compared inferences based on both original author of the commit and the committer, and found that they were virtually identical; regarding threat (ii), we validated the organization charts with the studied organization, and found them to be reasonably accurate and up-to-date; regarding threat (iii), the organization did not use automated merge tools and was not rebasing (i.e., re-writing) commits to any large extent.

The Git logs contained a small number of authors that could not be associated with a team.
As this concerned only a few data points (less than 2\% of the data), we modeled these as belonging to a separate ``Unknown'' team.
Although not correct from an organizational point of view, this allowed these data points to influence the population average, which would not have been possible if these data points had been excluded from the analysis. 

\anolrevised{We based our model on the detection of exact clones; Type~1 as defined by \citet{bellon2007} and \citet{koschke2007survey}.
We used the default configuration of \textsf{SonarQube} to detect clones.
Evaluating how the model performs for the identification of Type~2 (renamed/parameterized), Type~3 (near miss), and Type~4 (semantic) clones remains a subject for future studies. 
}

\textbf{Internal validity} deals with whether there might be other, non-studied factors that could explain some of the findings.

One complicating factor is that the study took place during the Covid-19 remote work period, where developers in their daily work had to work and collaborate remotely.
This possibly impacted both the onboarding of new teams and team members, as well as intra- and inter-team collaboration.
However, all the studied teams were operating under the same remote-work rules, so the same confounding factor applies to all teams in this study.
Still, teams might react differently to the remote-work mandate, but we have to defer this factor to a future study.
We found during the focus groups that even with the remote work mandate lifted, the studied organization was continuing to use a hybrid work mode, where teams would come to the office one or two days per week, and work the other days from the home office.

We focused our qualitative work on the core teams, with deep knowledge of the product.
This was partly for availability reasons, and partly because these teams were the most experienced, and had members that had been part of the product for a long time.
The same method could have been used to study less contributing teams, but then considerable effort would have to be spent to find the people who were part of these teams during the study period.

\textbf{External validity} concerns to what extent it is possible to generalize the findings, and to what extent the findings are of interest to other people outside of the investigated case.

This paper contains data from a particular system, developed by a particular organization, which limits generalizability.
However, as \citet{flyvbjerg2006} states, cases play an important role in human learning, and it is, in fact, possible to learn from a single case.

We have tried to describe characteristics that might enable others to judge whether our findings apply to other systems, but we cannot claim generalizability across all possible systems or organizations.
To aid this judgment, we provide the full anonymized data set, including all models and graphs described in the paper, and invite others to replicate the study in other contexts.

The constructs in our DAG and model are all generic and are not specific to the studied organization. 
Hence, it should be straightforward to use the same model in other contexts, to see whether the findings can be replicated.

\textbf{Reliability} concerns whether or not the data and analysis are dependent on the specific researchers.

Most of the data in this article are collected from quantitative sources, and processed and visualized using standard statistical tools.
We deliberately chose industry-standard tools (\textsf{git}, \textsf{SonarQube}) for data collection, to avoid bias specific to particular tools.
We provide a replication package, including the full anonymized data set, that can be used to replicate our Bayesian model-building process.

Interpretation of the processed data runs the risk of introducing reliability threats.
We strove to reduce these threats by interacting with the studied organization via mail and by setting up focus groups to discuss our findings. 
We elicited feedback from four teams, including architects, one line manager, and three core development teams.
Anonymized transcripts and codes are available upon request.

\section{Conclusions and Further Work}

\anolrevised{
This paper introduces a model and replication package ~\citep{epkanol_2024_11357296} designed to help visualize and understand how the team behavior related to introducing code clones varies across different components. The model is based on an industrial case study and has been validated through five focus-group sessions involving four teams.
}

The general pattern is that the number of introduced clones is related to both added and removed lines (i.e., the change size), and the existing complexity and number of duplicates,  following the \textit{Broken Window Theory}, as described by~\citet{hunt2000pragmatic}.

However, in some components, some teams are not affected by the existing complexity in the file.
Furthermore, teams that have chosen ownership of a component appear to introduce fewer duplicates, relative to how they behave in other components.
This suggests that (self-selected) ownership does matter in getting the teams to introduce fewer code clones.

An important factor to consider when analyzing data is \textit{why} teams behave as they do. 
In the study, one team that \textit{stood out}, relative to the other teams, was tasked to rearchitect the main business flow of the application, and this is a probable explanation for their behavior.
Thus, the visualizations are important, but the \textit{insights} that the visualizations lead to are probably more important (e.g., the importance of removing old code, once the new architecture is in place).

All interviewed teams stated that the visualizations of the clone introduction probabilities were useful for understanding clone introduction behavior across teams and repositories.
However, most teams also felt that the model should be complemented with a corresponding visualization of the amount of removed clones.
We developed the simplest possible model in this regard and found no significant outliers on a team-level stratification.
In most repositories, most teams were acting similarly regarding clone removals.

This suggests that these two models could make a useful addition to a toolbox to govern the commons, according to the principles outlined by~\citet*{dietz2003struggle}.
We intend to explore the clone removal model in follow-up studies and invite others to do the same.

\anolrevised{
Understanding how much historical data to incorporate into models like these, and how often to resample them remains to be studied in follow-up studies.

In addition, replications in other organizations are needed to understand whether the observed behaviors differ in other contexts, and whether the models can be useful when analyzing other units of analysis (such as geographical sites or sub-organizations).
}

\backmatter
\bmhead{Acknowledgments}
This research was supported by the Knowledge Foundation through the KKS Profile project SERT 2018/010 at Blekinge Institute of Technology, Sweden.

\bmhead{Data Availibility}
The collected, anonymized, dataset, plus the resulting converged model, are available as an executable replication package on Zenodo and Github~\citep{epkanol_2024_11357296}.
The scripts used to build the dataset from SonarQube analyses are available upon request.

\section*{Declarations}
\bmhead{Conflict of Interest}

Mr Sundelin declares that he has been employed by Ericsson AB and Ericsson Mobile Financial Services AB during the writing of this study, but neither company nor any other would gain or lose financially from reporting this research.
The other authors declare that they have no known competing financial interests or personal relationships that could have appeared to influence the work reported in this paper.


\bibliography{references}

\end{document}